\documentclass[preprint,5p,twocolumn]{elsarticle}

\usepackage{graphicx}
\usepackage{booktabs}
\usepackage{amsmath}
\usepackage{hyperref}
\usepackage{float}
\usepackage{subcaption}
\usepackage{cleveref}
\usepackage{makecell}
\usepackage{subcaption}
\usepackage{natbib}

\journal{Nuclear Instruments and Methods in Physics Research Section A}

\biboptions{sort&compress}

\begin{document}
\begin{frontmatter}

\title{A non-invasive ultra-thin luminophore foil detector system for secondary beam monitoring}

\author[PSI,ETHZ]{F. Berg\fnref{Schaer}}
\author[BINP,NSTU,NSU]{D.N. Grigoriev}
\author[PSI,ETHZ]{Z. Hodge\fnref{UniBE}}
\author[PSI]{P.-R. Kettle\corref{thecorrespondingauthor}}
\ead{peter-raymond.kettle@psi.ch}
\author[BINP,NSU]{E. A. Kozyrev}
\author[BINP]{A.G. Lemzyakov}
\author[BINP]{A.V. Petrozhitsky}
\author[BINP,NSU]{A. Popov}

\address[PSI]{Paul Scherrer Institut PSI, 5232 Villigen-PSI, Switzerland}
\address[ETHZ]{Swiss Federal Institute of Technology ETH, 8093 Z\"{u}rich, Switzerland}
\address[BINP]{Budker Institute of Nuclear Physics of Siberian Branch of Russian Academy of Sciences, 630090 Novosibirsk, Russia}
\address[NSTU]{Novosibirsk State Technical University, 630092 Novosibirsk, Russia}
\address[NSU]{Novosibirsk State University, 630090 Novosibirsk, Russia}

\fntext[Schaer]{Present address: Schaer Proton AG, 8416 Flaach, Switzerland}
\fntext[UniBE]{Present address: University of Bern, 3012 Bern, Switzerland}
\cortext[thecorrespondingauthor]{Corresponding author}

\begin{abstract}
High-intensity secondary beams play a vital role in today's particle physics and materials science research and require suitable detection techniques to adjust beam characteristics to optimally match experimental conditions.
To this end we have developed a non-invasive, ultra-thin, CsI(Tl) luminophore foil detector system, based on CCD-imaging.
We have used this to quantify the beam characteristics of an intensity-frontier surface muon beam used for next-generation charged lepton-flavour violation (cLFV) search experiments at the Paul Scherrer Institut (PSI) and to assess the possible use for a future High-intensity Muon Beam (HiMB-project), currently under study at PSI.
An overview of the production and intrinsic characteristics of such foils is given and their application in a high-intensity beam environment.
\end{abstract}


\end{frontmatter}


\section{Introduction}\label{sec:1}
Inorganic luminophore based screens made of ZnS have been known since the turn of the 20th Century when first used as single particle detectors \cite{Crookes1903} or for the pioneering work on atomic structure \cite{Geiger1913}. 
Since then the luminescent properties of luminophores have been developed from both organic and inorganic compounds and tailored to a multitude of applications and subjects from high-energy particle physics experiments through x-ray tomography and medical diagnostics to accelerator beam diagnostics as well as industrial applications such as new display technologies.

High-intensity secondary beams of muons and pions are in ever increasing demand to match the required intensity needs of a spectrum of fundamental experiments both in the low-energy precision sector: cLFV, muonic-atom/muonium spectroscopy, muon anomalous magnetic moment/electric dipole moment (g-2/EDM) and at the high-energy scale, such as the front-end sources for a neutrino factory or muon collider or the high-energy version muon g-2 measurement. 

In order to achieve the sensitivity goal of such experiments, it is essential to be able to reliably check the beam optics and have the ability to tune the beams to maximize intensity, while minimizing beam-correlated background and maintaining a good achromaticity at the final focus for stopped beams. 
This ideally requires non-invasive, in vacuum, detectors capable of working at high intensities and being sufficiently radiation hard.

The most intense continuous surface muon beams in the world exist at PSI, with muon yields in excess of 10$^{8}$ muons/s \cite{PSIHandbook}. 
The PiE5 channel coupled to either the MEG or Compact Muon Beam Line (CMBL) feeds the two cLFV search experiments MEG~II \cite{Baldini2018} and phase I of Mu3e \cite{Blondel2013}, which both require running at the maximum intensity and hence require beam diagnostics during tuning and ideally non-invasive diagnostics during data-taking. The former was generally achieved using an automated remote 2-D beam scanning pill scintillation detector system. 
Although a very reliable and stable technique, able to handle rates $\mathcal{O}$(10$^{9}$), it is however an invasive and somewhat time-consuming method. 
In order to improve on this, a non-invasive, in-situ vacuum measurement was sought capable of handling a high-rate environment and without corrections required for multiple scattering in beam windows or air, as with the beam scanner system. 
For low-energy surface muon beams of 28~MeV/c with a kinetic energy of 3647~keV, close to the kinematic edge of stopped pion decay, very thin foils at the sub-10-micron level are required, if not used as a degrader. 
This is necessary in order to minimize multiple scattering while maintaining sufficient light-yield for image capture, since the mean range is of the order of 480~$\mu$m in CsI. 
To this purpose we have developed a luminophore foil of CsI(Tl) of a few microns in thickness on a polyester (PET) substrate of 3 microns, produced by thermal vapour deposition.

Although the use of CCD imaging of CsI(Tl) plates/screens/foils has been known for a while, its usage has been mainly in the areas of X-ray imaging/radiography \cite{Nagarkar1998,Cha2009} or with primary ion beams of higher current density involving beam profiling with either thicker layers of CsI or thicker substrates \cite{Cosentino2003,Harasimowicz2010,Re2005}. 
Hence, primarily beam destructive measurements. 
Secondary beams such as pions and muons, even at the intensity frontier, distinguish themselves from their primary production counterparts by their larger emittances and smaller current densities. 
Secondary beam monitoring techniques employed at some of the dedicated pion and muon beam facilities around the world such as RIKEN-RAL in the UK, J-PARC MUSE facility in Japan or PSI in Switzerland, use extractable, destructive imaging techniques of scintillation plates \cite{Eaton1994,Lord2010,Ito2014} placed in the beam. 
Other less invasive techniques such as scintillating fibre arrays are also used \cite{Stoykov2005,Strasser2010,Ripiccini2016}.
In this paper we concentrate on the use of ultra-thin luminophore detection foils capable of detecting a low-energy continuous particle beam with beam-spot current densities ranging from $\sim$ (0.3-30)~fA/mm$^{2}$, corresponding to (10$^{7}$-10$^{9}$) particles/s.
In \Cref{sec:2} we give a brief overview of the foil production, the CCD imaging systems used and general image analysis procedure, as well as the intrinsic characteristics of such foils, while in \Cref{sec:3} we demonstrate their usefulness in characterizing the beam properties of a high-intensity surface muon beam and compare the results to our standard 2-D remote scanning pill scintillator system. 
In \Cref{sec:4} we briefly look to the future.

\section{Luminophore foil production, CCD imaging and foil intrinsic characteristics}\label{sec:2}
CsI(Tl) is a well-known scintillation medium, with a high light yield of $\sim$ 54 thousand photons per MeV energy deposition \cite{SGCsI}. 
It also has an optimal emission spectrum, well matched to the sensitivity of most common CCD visible light sensors (see below) and hence is a favourable candidate as a luminophore foil beam monitor.

\subsection{Foil production}
The thermal vapour deposition technique we used is described elsewhere \cite{Kozyrev2016} in the context of later developed luminophores specifically for X-ray imaging, so here we provide only a brief overview together with the differences.
A mixture of CsI and Tl (doping concentration $\sim$0.08~mole\%) was maintained at a temperature of 680~$^{\circ}$C in a tantalum boat placed some 65 cm below the rotating, frame-mounted ultra-thin PET substrate. 
The 3~$\mu$m thick and 130 mm diameter substrate was maintained at 25~$^{\circ}$C and subjected to a pressure of 5-10$^{-3}$~Pa during the process. 
The deposition rate was kept low (17$\pm$2~\AA/s) in order to achieve an even coverage. 
The CsI(Tl) layer structure, when analyzed by a scanning electron microscope (SEM), was found to consist of grains between 2-5~$\mu$m in size \cite{Kozyrev2016}. 

\begin{figure}[H]
   \centering
   \includegraphics[width=\linewidth]{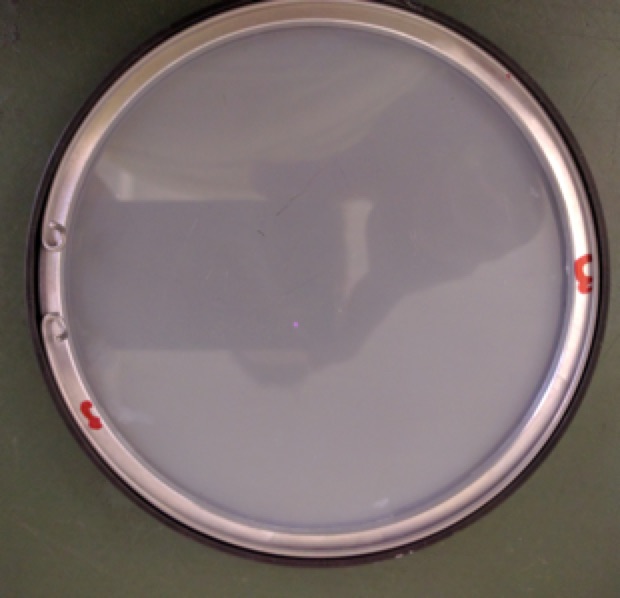}
   \caption{Example of a luminophore foil, showing Foil 3 with an active diameter of 125 mm.}
   \label{fig:1}
\end{figure}

Of the four foils employed in these measurements see \Cref{tab:1}, one (foil~4) had a 100~nm layer of Al deposited on the PET substrate prior to vapour deposition.
This was intended to act as a mirror, however, the CsI layer here was only 3~$\mu$m thick, whereas the foils 1-3 had layers of between 5.0-5.2~$\mu$m thickness.
None of the foils had a special surface coating added as mentioned in \cite{Kozyrev2016}.
An example of such a frame-mounted foil is shown in \Cref{fig:1}.

\begin{table}[!h]
  \centering
  \caption{Luminophore foil characteristics used in the measurements.}
	\begin{tabular*}{\linewidth}{@{\extracolsep{\fill}} l c c c @{}}
	\toprule
	Foil & \thead{\small CsI layer \\ {[$\mu$m]}} & \thead{\small Al layer \\ {[nm]}} & \thead{\small Active dia. \\ {[mm]}} \\
	\midrule
	Foil 1	& 5.0 		& - 	& 125 \\
	Foil 2	& 5.2		& - 	& 125 \\
	Foil 3	& 5.1		& - 	& 125 \\
	Foil 4	& 3.0		& 100	& 125 \\
	\bottomrule
 	\end{tabular*}
  \label{tab:1}
\end{table}

\subsection{CCD Imaging and image analysis overview}\label{sec:2.2}
\begin{table*}[htp]
	\centering
	\caption{Main technical specifications of the IDS, ORCA and QSI camera systems taken from \cite{IDS,ORCA,QSI}.}
	\begin{tabular*}{\linewidth}{@{\extracolsep{\fill}} l c c c c c @{}}
		\toprule
		Camera	& \thead{\small Resolution (Sensor) \\ {[pixels]}}	& \thead{\small Pixel Size \\ {[$\mu$m]}}	& \thead{\small Full Well Capacity \\ {[e]}}	& Bit Depth	& Peak QE \\ 
		\midrule
		\begin{tabular}{@{}l@{}} IDS \\ UI-2220SE \end{tabular}	& \begin{tabular}{@{}c@{}} 768$\times$576 \\ (Sony ICX415AL) \end{tabular}	& 8.3$\times$8.3	& 15300	& 8	& 39\% @~533~nm \\ 
		\midrule
		\begin{tabular}{@{}l@{}} Hamamatsu \\ ORCA Flash 4.0 V2 \\ C11440-22C \end{tabular} 	& \begin{tabular}{@{}c@{}} 2048$\times$2048 \\ (Sci. CMOS Sen. typ FL-400) \end{tabular}	& 6.5$\times$6.5	& 30000	& 16	& 73\% @~560~nm \\ 
		\midrule
		\begin{tabular}{@{}l@{}} QSI \\ RS9.2S \end{tabular}	& \begin{tabular}{@{}c@{}} 3388$\times$2712 \\ (Sony ICX814)	 \end{tabular} & 3.69$\times$3.69	& 18000	& 16	& 77 \% @~560~nm \\
		\bottomrule
	\end{tabular*}
	\label{tab:2}
\end{table*}
In total three different types of Camera system were used during the development of the beam monitor described here.
The relevant technical specifications for each system are shown in \Cref{tab:2}.
The IDS UI-2220SE is a non-cooled, low resolution (0.44~Mpixels) CCD camera with an 8-bit depth, while the 16-bit high resolution (4~Mpixels) Hamamatsu ORCA Flash 4.0 is a thermoelectric/liquid cooled CMOS camera with a much higher dynamic range than the IDS.
Both of these cameras were used extensively for the muon beam measurements.
The 16-bit QSI RS-9.2S camera with its high resolution (9.2~Mpixels) was procured for the final beam line installation and specifically used for the luminophore characterization, as this thermoelectric/liquid cooled CCD with its particularly high dynamic range shows, when cooled, a superior dark current to the other cameras.
The main lens used for the majority of measurements was a Canon Telephoto lens EF-S 18-200~mm f/3.5-5.6~IS, which served to meet the varying setup conditions required during the testing, in and out of the beam.

Analysis of the CCD images was common to all cameras and involved taking a series of typically five beam images, as well as an equivalent set of background images.
In the case of beam imaging the background was taken with the channel beam-blocker closed and for the same exposure time as the beam image, so enabling the subtraction of any stray light from other sources, as well as the inherent thermal noise of the sensor.
Furthermore, calibration images of the foil with its calibration-grid (see \Cref{fig:21}) illuminated using a UV-LED were also taken and served two-fold: firstly, as means of calibrating the image pixel-to-mm ratio and secondly to perform any image perspective transformations required when viewing the foil off-axis.
After each image has undergone its perspective transformation \cite{Bradski2008} using the transformation matrix derived from the calibration image, the individual pixels of each image of the set are summed and normalized to a nominal 2~mA proton beam intensity (the measured proton beam intensity during the exposure is proportional to the secondary particle rate), the exposure time being the same for all images.
The same procedure is adopted for the background images, each set, beam and background are then averaged, and the background average subtracted from the beam average image yielding a background subtracted, normalized  fully calibrated beam image with its associated pixel variances.
From this image a region-of-interest (ROI) is selected, excluding the foil frame and a 2-D correlated Gaussian function is fitted to the image.
This then yields the beam centroid position and the transverse widths and correlation, while the secondary beam intensity is obtained from an initial one-off independently measured secondary particle to proton intensity ratio.

\subsection{Luminophore intrinsic properties}
The characterization of the luminophore foil involved measuring the relevant intrinsic properties, such as one of the main spectral attributes, the emission spectrum, as well as assessing the homogeneity of the CsI(Tl) layer.
Finally, a study of the radiation hardness of such a foil was made using the Budker Institute's (BINP) ILU-10 electron accelerator at Novosibirsk.

Both the spectral and homogeneity measurements used a monochromatic UV-source centred at 285~nm.
The measurements were performed in a light-tight box with diffuse black, non-reflecting surfaces.
The UV-LED was shone at the CsI(Tl) side of the foil, allowing a partial reflection.
For the emission spectrum measurements, the emitted light was analyzed by an Ocean Flame-S-XR1-ES spectrometer, sensitive between 200-1025~nm.
In the case of the homogeneity measurement the light yield (LY) was analyzed using the CCD-camera with and without a UV-filter to cut-out the UV-LED emission below approximately 390~nm.  

\begin{figure}[H]
   \centering
   \includegraphics[width=\linewidth]{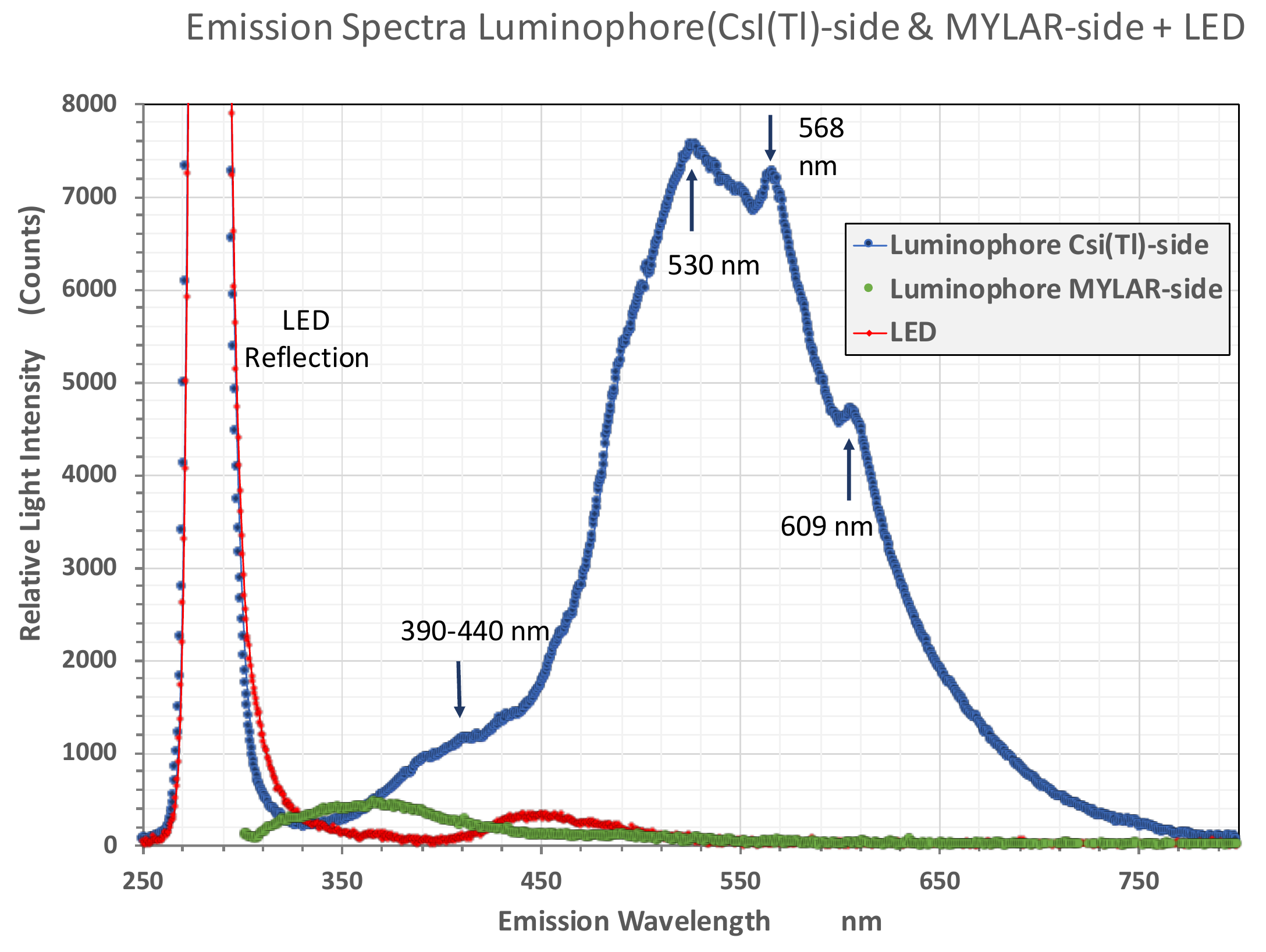}
   \caption{Measured total emission spectrum of the luminophore CsI(Tl) foil and its components: PET, CsI(Tl) and the UV-LED itself. See text for details.}
   \label{fig:2}
\end{figure}

The emission spectrum of all three components, the UV excitation source, a Thorlabs LED-285W, the PET substrate (Mylar\textsuperscript{\textregistered})~and the CsI(Tl) layer were measured and are shown in \Cref{fig:2}.
The total luminophore spectrum, including all three components, is shown by the blue points, while the PET emission spectrum obtained by exciting and measuring the substrate side of the foil is shown by the green  points.
This was cross-checked by comparing it to a separately measured pure Mylar\textsuperscript{\textregistered}~foil spectrum, which showed the same characteristics.
Also shown, in red, is the emission spectrum of the UV-LED, normalized to the reflection in the total spectrum.

The characteristic visible and UV emission bands from the Tl-activator \cite{Gridin2014,Lagu1961} can clearly be seen at 530-570~nm, while only a shoulder can just be discerned from the other visible band at $\sim$ 500~nm.
The UV-band around 390-440~nm can also be seen, though also as a shoulder, due to both the narrow absorption band width \cite{Gridin2014} and the almost monochromatic excitation source centred around 285~nm.
An excitation peak around 609~nm can also be seen in the figure, which has also been addressed in the literature \cite{Gridin2014}. 

\begin{figure}[H]
   \centering
   \includegraphics[width=\linewidth]{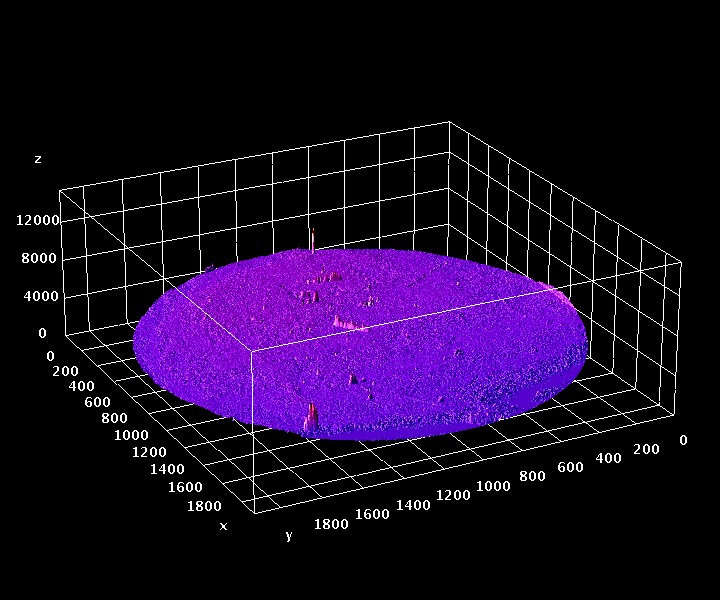}
   \caption{Colour density plot of LY from foil 3 excited by the UV-LED. The CCD-camera has a UV-filter, cutting-off wavelengths below 390~nm.}
   \label{fig:3}
\end{figure}

\begin{figure}[H]
   \centering
   \includegraphics[width=\linewidth]{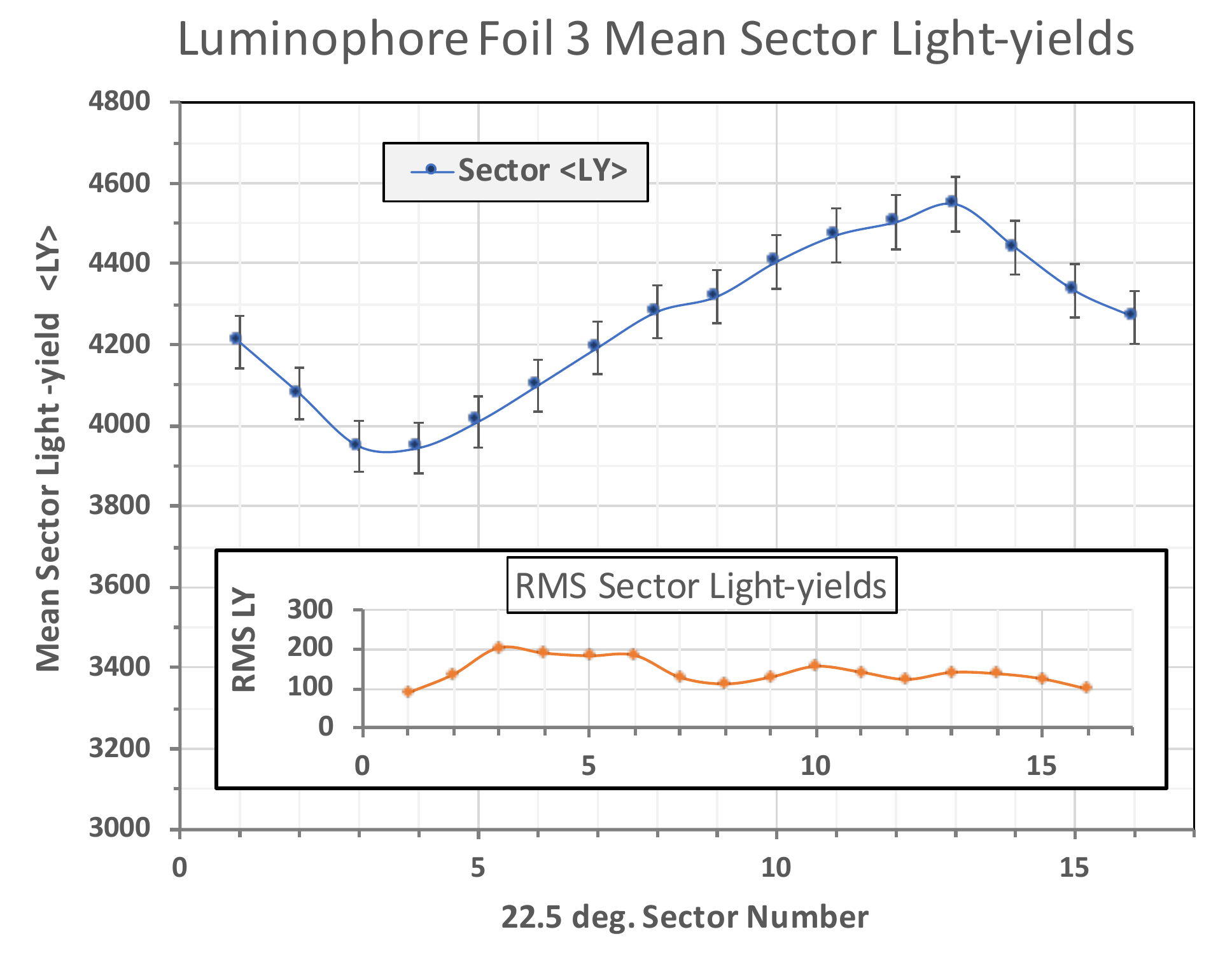}
   \caption{Plot of mean azimuthal sector LYs for the total ROI corresponding to 16$\times$22.5$^\circ$ sectors. The insert shows the corresponding rms values per sector. The variation of the LY from the average is less than 8\%.}
   \label{fig:4}
\end{figure}

The Light-yield (LY) homogeneity of the foils was sampled by means of measurements using the UV-LED with its monochromatic excitation wavelength centred at 285~nm.
The LY of foil 3 was measured using the QSI camera with a UV-filter placed on the lens to cut wavelengths below 390~nm, hence only being sensitive to the emission spectrum of CsI(Tl), whereas foil 1 was measured in situ in the beam line vacuum system with a remote UV-LED placed upstream in vacuum.
The LY for foil 1 was measured using the IDS camera system without a UV-filter.
Again, the standard background subtraction analysis procedure described in \Cref{sec:2.2} was adopted yielding an averaged, background subtracted excitation image. 
The LY-distribution for foil 1 was analyzed by segmenting the ROI, which excludes the foil frame and edge, into 5x5~mm$^{2}$ squares which are fully contained. 
Each region's summed pixel value is then placed in a 2D histogram. 
The variation in LY homogeneity is taken from the maximum difference of LY between squares, which corresponds to 13\% for foil 1. 
An independent set of measurements using the QSI-camera with UV-filter were undertaken with foil 3 in the light-tight box. 
Here the UV-light source was place such as to indirectly illuminate the foil allowing a more homogeneous illumination.
The ROI was then segmented into 16 equal azimuthal sectors each of 22.5$^\circ$.
The average and rms deviation for each sector were then compared.
A 2D light-yield intensity plot of the full ROI is shown in \Cref{fig:3}, where several surface defects are clearly visible.
A plot of the average sector LYs and the corresponding rms-values are shown in \Cref{fig:4}.
The maximum variation of the LYs corresponds to 15.3$\pm$2.3\% giving a variation of less than 8\% from the average LY, consistent with foil~1.
However, this is considered an upper limit on the LY homogeneity as the evident left-right asymmetry seen in \Cref{fig:4} could still have a partial source reflection dependence.

\begin{figure}[H]
   \centering
   \includegraphics[width=\linewidth]{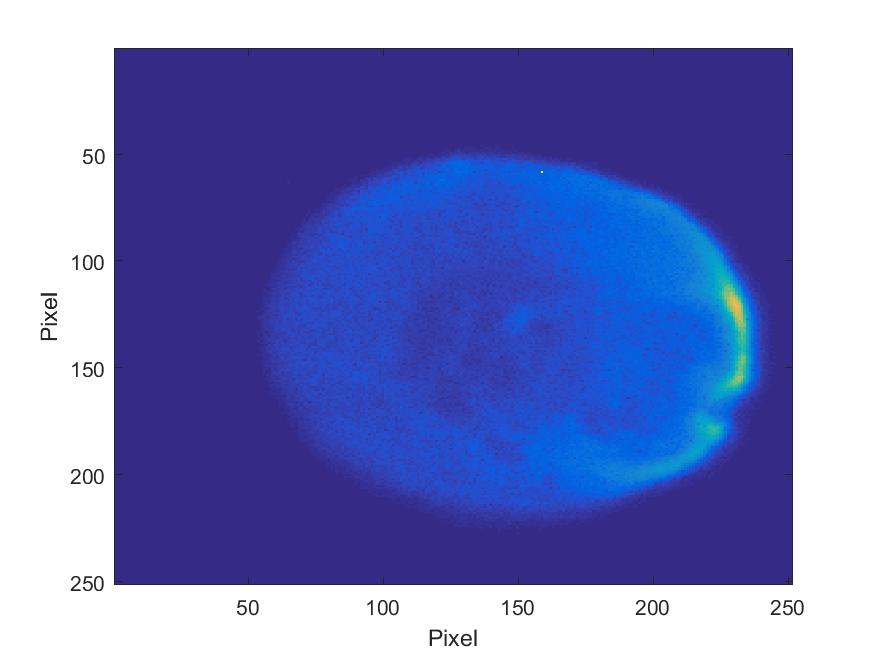}
   \caption{Colour density plot of the light-yield versus the transverse coordinates of the foil in pixels. The luminophore test foil views a $^{226}$Ra source, post irradiation. The colour density variation in the picture is due to the inhomogeneity of the $^{226}$Ra source itself.}
   \label{fig:5}
\end{figure}

\begin{figure}[H]
  \centering
  \begin{subfigure}[t]{\columnwidth}
  \includegraphics[width=\linewidth]{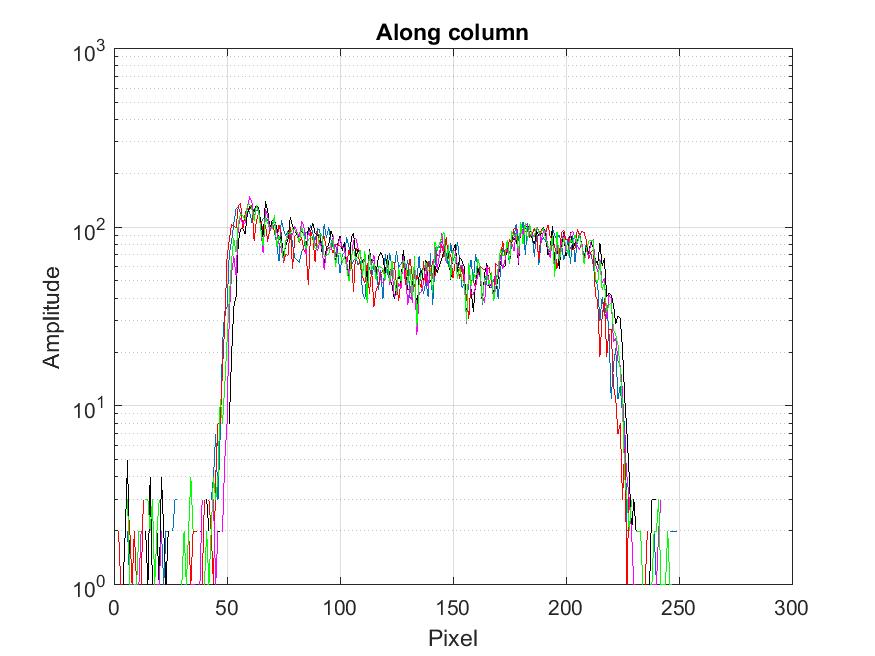}
  \caption{}
  \label{fig:6a}
  \end{subfigure}
~
  \begin{subfigure}[t]{\columnwidth}
  \includegraphics[width=\linewidth]{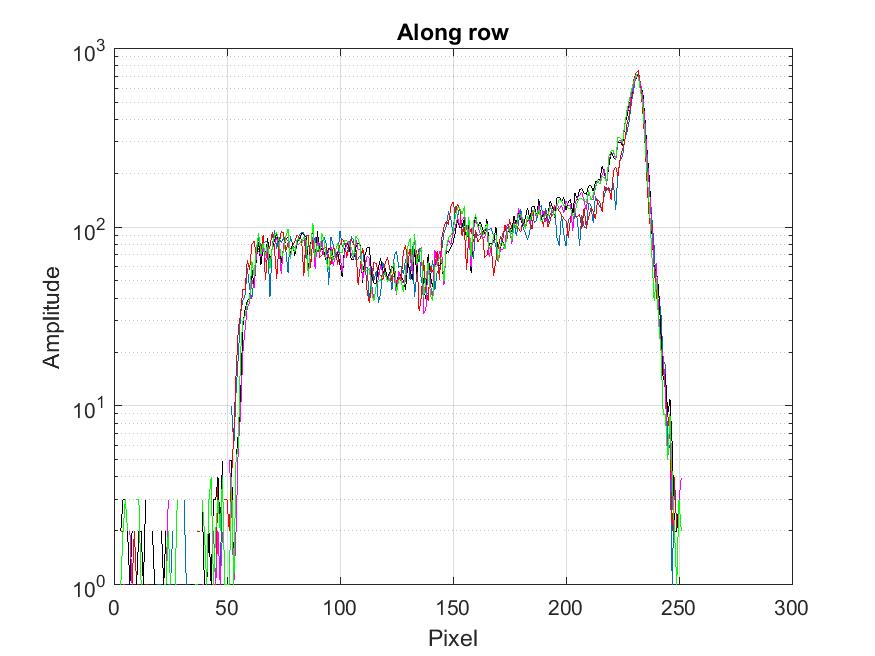}
  \label{fig:6b}
  \caption{}
  \end{subfigure}
  \caption{Log plot of the superimposed column (top) and row (bottom) LY, showing the influence of the alignment precision.}
  \label{fig:6}
\end{figure}

The final intrinsic property studied was the effect of ionizing radiation, in the form of fractionated doses from a low-energy, 5~MeV electron beam, on the LY of such foils.
This was carried out with an ILU-10 industrial accelerator at BINP using a small foil of 38~mm diameter. 
This underwent a fractionated irradiation programme in 15~kGy steps, up to a total dose of 75~kGy. 
This dose is approximately 30\% higher than the estimated total central 1$\sigma$ muon spot dose expected from a 1-year running period of the MEG~II experiment at PSI. 
The estimated energy deposition per 28~MeV/c muon in the CsI(Tl) layer is $\sim$ 21~keV leading to a 1$\sigma$-beam spot dose rate of $\sim$ 0.46~kGy/day. 
The effect of the fractionated dose on the LY was studied after every 15~kGy increase by measuring the LY of the foil exposed to a $^{226}$Ra source. 
A picture of the foil after one such fractionation is shown in \Cref{fig:5}.
The total LY is obtained from the integral of the pixels after background subtraction.
The non-uniformity in LY seen in \Cref{fig:5} is due to the inhomogeneity of the source itself.
The systematic uncertainties associated with the integrated dose measurements are two-fold: the first is the uncertainty on the absolute dose, which is estimated to be $\sim$ 30\%, though the relative variation between fractionations is much smaller; the second and significant contribution comes from the reproducibility of the alignment of source, foil and camera for each measurement, which was determined by comparing the LY from the same selected central pixel row and column.
This is shown in \Cref{fig:6} where the pixel rows and columns for each image, taken after each fractionation measurement, are superimposed.
The estimated uncertainty on each measurement is 0.7\%.

\begin{figure}[H]
   \centering
   \includegraphics[width=\linewidth]{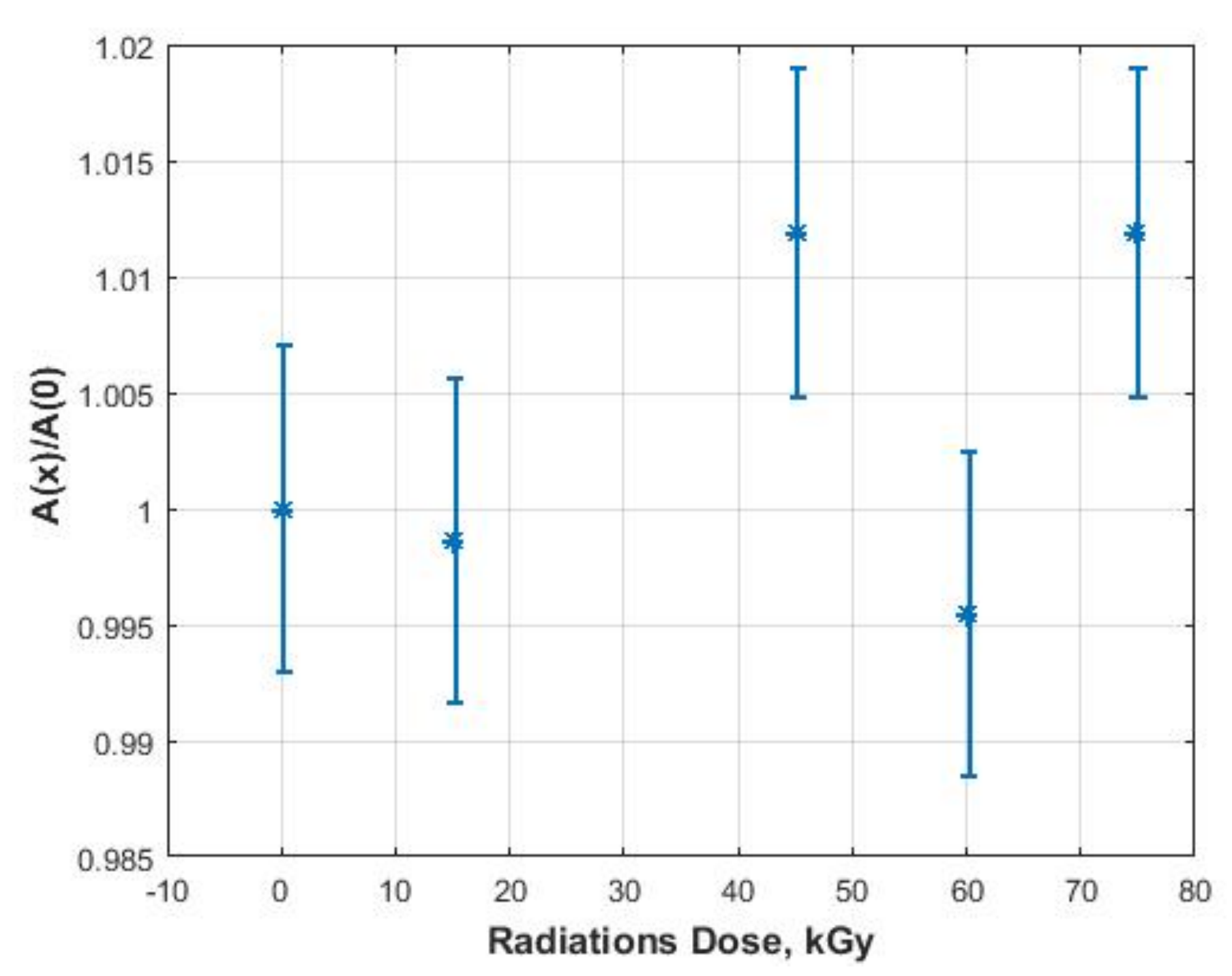}
   \caption{Relative LY-response of the luminophore foil to fractionated 15~kGy doses from a 5~MeV electron beam.}
   \label{fig:7}
\end{figure}

\Cref{fig:7} shows the final relative LY-change versus integrated dose, where $A(0)$ is the integral of the LY over the image before irradiation and $A(x)$ is the LY after the fractionated dose $x$. 
Statistically, the LY-ratio $A(x)/A(0)$ is compatible with 1.0 up to the maximum dose induced of 75 kGy.

\section{Beam characterization using the luminophore foil}\label{sec:3}
\begin{figure}[H]
   \centering
   \includegraphics[width=\linewidth]{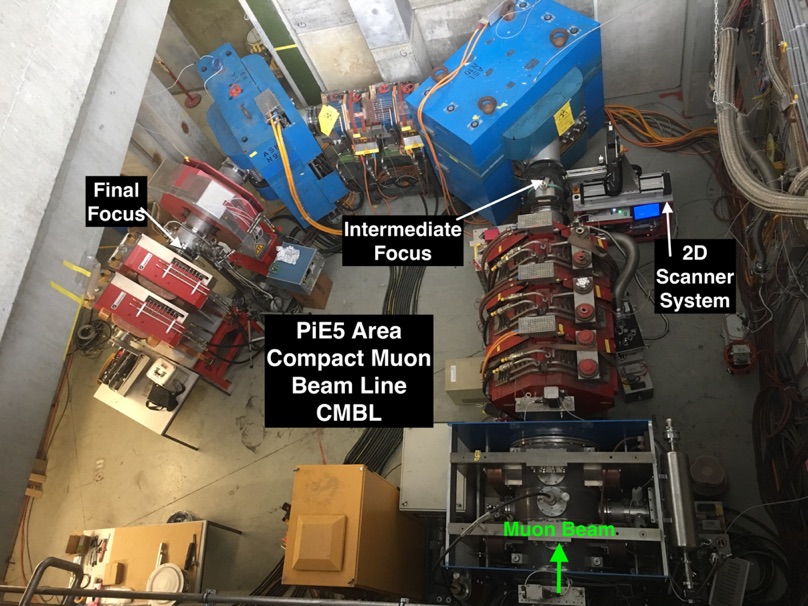}
   \caption{CMBL beam line configuration in the PiE5 area at PSI with two extra quadrulpole magnets beyond the final focus, in place of the large Mu3e solenoidal magnet. Luminophore measurements were undertaken at the intermediate and final foci. The 2-D pill scanner system can be seen placed at the intermediate focus.}
   \label{fig:8}
\end{figure}

The muon beam measurements described here were undertaken at the MEG and CMBL beam lines coupled to the PiE5 channel at PSI, a more detailed description of the measurement environment is given in \cite{Baldini2018,Blondel2013}, as well as a brief overview of the first usage of the luminophore foil \cite{Baldini2018}. 
The beam line setup in its CMBL configuration is shown in \Cref{fig:8}. 
Here we give a detailed view of the characteristics of the luminophore system comparing it to measurements made with our standard pill scintillation scanning system, demonstrating that this new luminophore system can be used to characterize one of the highest intensity continuous surface muon beams in the world.

\subsection{Experimental setup}
\begin{figure}[H]
   \centering
   \includegraphics[width=\linewidth]{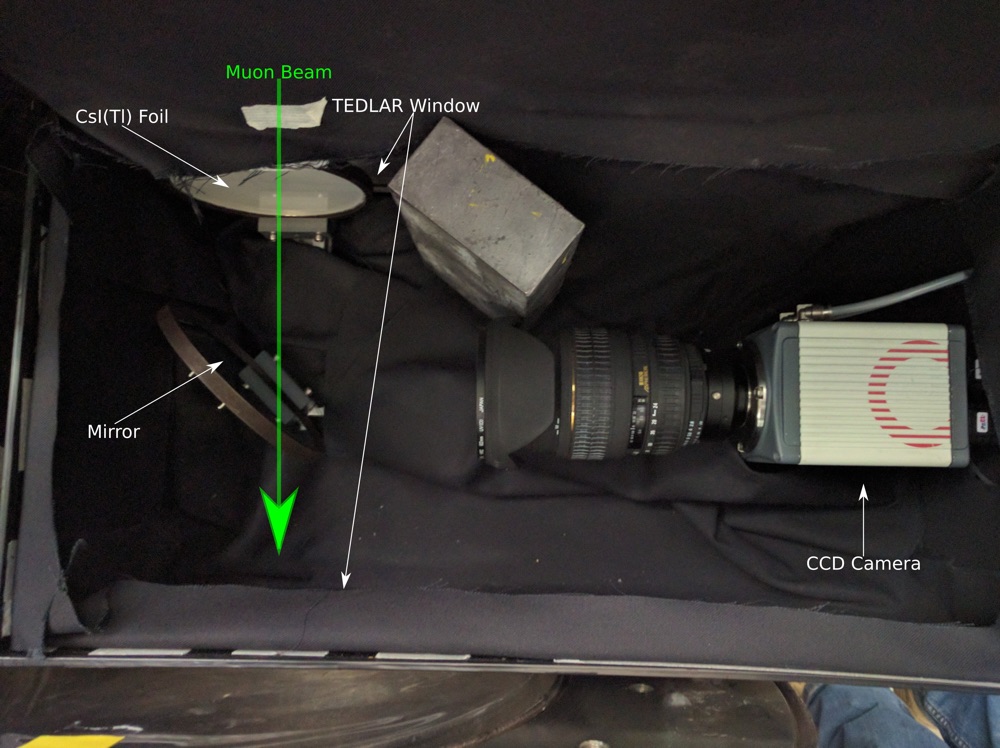}
   \caption{Light-tight box setup used for beam measurements in air.}
   \label{fig:9}
\end{figure}

Two basic setups were used for the measurements, one a setup in air using a light-tight box which was placed immediately downstream of the beam line vacuum window, this had either a 20~$\mu$m Aluminium or a 50~$\mu$m Tedlar\textsuperscript{\textregistered}~(polyvinyl fluoride PVF) light-tight entrance and exit window for the muon beam. 
The luminophore was placed at 90$^\circ$ to the incoming beam, just behind the entrance window and was viewed under 90$^\circ$ to the beam-axis by means of a 3~$\mu$m thick aluminized PET foil mirror, placed at 45$^\circ$, and viewed by the CCD camera, allowing the beam to exit with minimal interaction. 
A picture of the setup is shown in \Cref{fig:9}.

\begin{figure}[H]
   \centering
   \includegraphics[width=\linewidth]{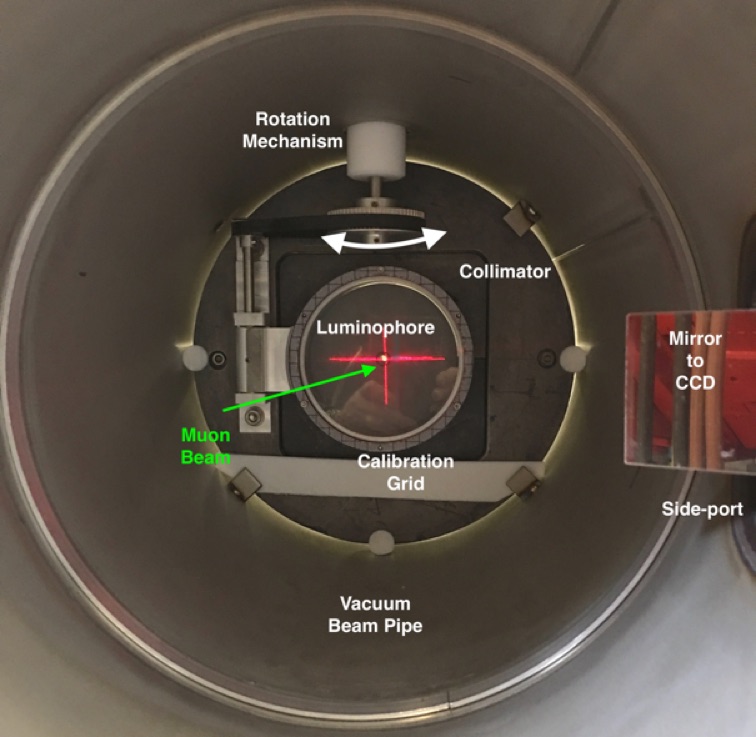}
   \caption{Vacuum measurement setup at the intermediate focus showing the collimator system with the rotatable luminophore foil holder. The foil with its calibration grid can be seen at the centre of the picture. The beam-spot is imaged via a CCD camera through a vacuum window in the side port of the beam pipe via a mirror.}
   \label{fig:10}
\end{figure}

The second setup in vacuum is now an integral part of the beam line, incorporated into the collimator system at the intermediate focus. 
This setup is shown in \Cref{fig:10}, a view of the inside of the vacuum beam pipe with the luminophore and its calibration-grid in place, covering the full 120~mm aperture of the collimator which is used to stop the deflected 28 MeV/c positron beam separated by the upstream Wien-filter from the undeflected surface muon beam. 
A rotation mechanism allows the luminophore to be moved in/out of the beam. 
The foil is viewed via a mirror support structure mounted upstream, in a side-port of the vacuum-pipe with a window interface to a CCD camera outside. 
For calibration purposes, a UV-LED is mounted far upstream in an upper port of the beam-pipe which is used to excite/illuminate both the grid and the foil.

Finally, before describing the actual measurements a brief overview of the comparative 2-D pill scintillator scanner measuring system is given as this is used as a comparison for the luminophore measurements. 

\begin{figure}[H]
   \centering
   \includegraphics[width=\linewidth]{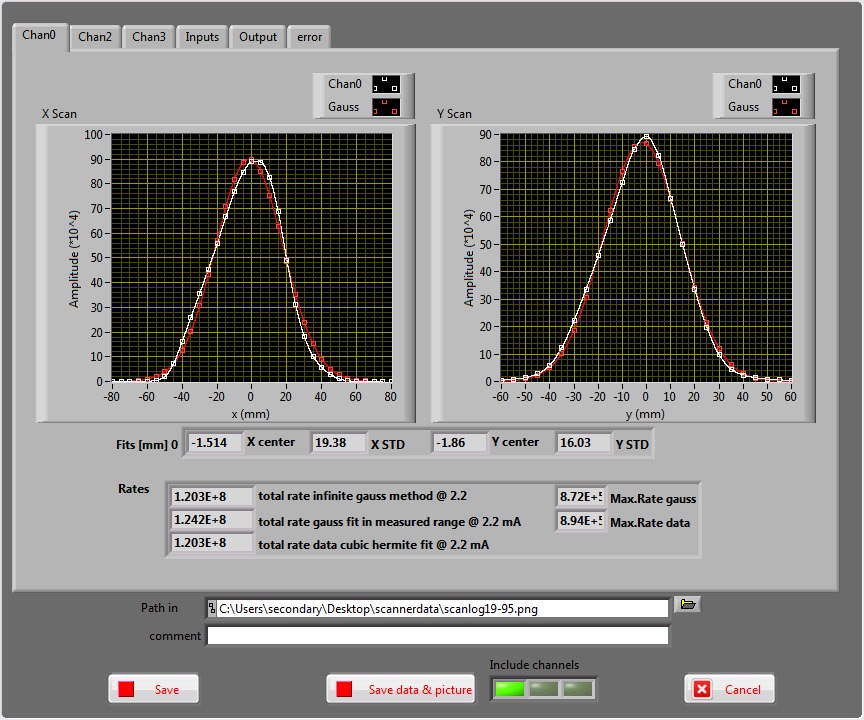}
   \caption{High-threshold cross-scan output showing the horizontally and vertically measured beam profiles with centroids and sigmas as well as the total intensity, from the various fit models.}
   \label{fig:11}
\end{figure}

The system is comprised of a remotely controlled 2-axis, sub-millimetre precision, stepping-motor driven scanning arm with a fixed-mounted Hamamatsu R9880U-110 miniature photomultiplier tube (PMT) and an NE102A pill-scintillator of 2~mm diameter and 2~mm length (sufficient to totally stop the surface muons). 
A 4-channel waveform digitizing board (DRS4 evaluation board \cite{Ritt2010}) plus fast electronics and a LabVIEW-based \cite{Elliott2007} acquisition system measures the threshold dependent particle rates (muon and positron) in the scintillator, normalizes them against the also measured proton beam intensity as well as yielding a particle time-of-flight measurement between the production target and the scintillator, from the particle timing with respect to the measured accelerator modulo 19.75~ns RF-structure. 
These measurements are repeated for different x- and y-positions of the scintillator, either in form of a so-called ``cross-scan" in which an initial rough scan automatically determines the true beam centroid and then performs a 5~mm step horizontal- and vertical-scan through the centroid. 
The acquisition then outputs a plot of the horizontal and vertical beam profiles together with a measure of the proton beam normalized particle intensity, obtained by  various fitting methods to the profiles. 
An example of such a cross-scan output is shown in \Cref{fig:11}. 
The other form of scan performed is a so-called 2-D ``raster scan", which involves taking measurements on a grid of scan positions (raster) with 5~mm spacing in both x and y. 
Again, the beam centroid and corresponding sigmas are determined though this time from a 2-D Gaussian fit to the beam spot. 
This type of scan can involve a total of $\sim$ 1400 single measurements taking some 2.5 hours, whereas a corresponding cross-scan takes about 10 mins.

\subsection{ Luminophore Measurements}
\begin{figure}[H]
  \centering
  \begin{subfigure}[b]{0.99\columnwidth}
  \includegraphics[width=\linewidth]{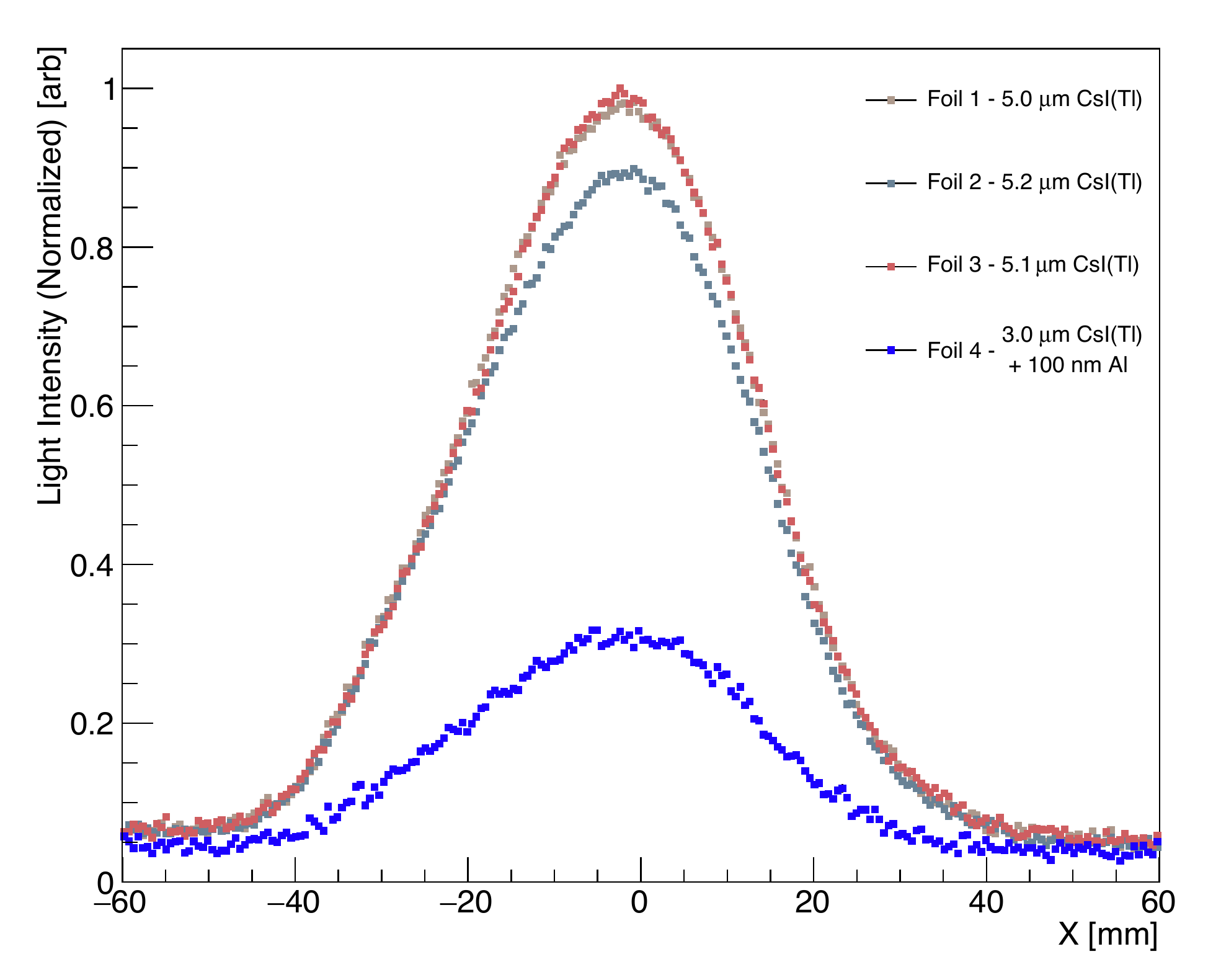}
  \caption{}
  \label{fig:12a}
  \end{subfigure}
\vfill
  \begin{subfigure}[b]{0.99\columnwidth}
  \includegraphics[width=\linewidth]{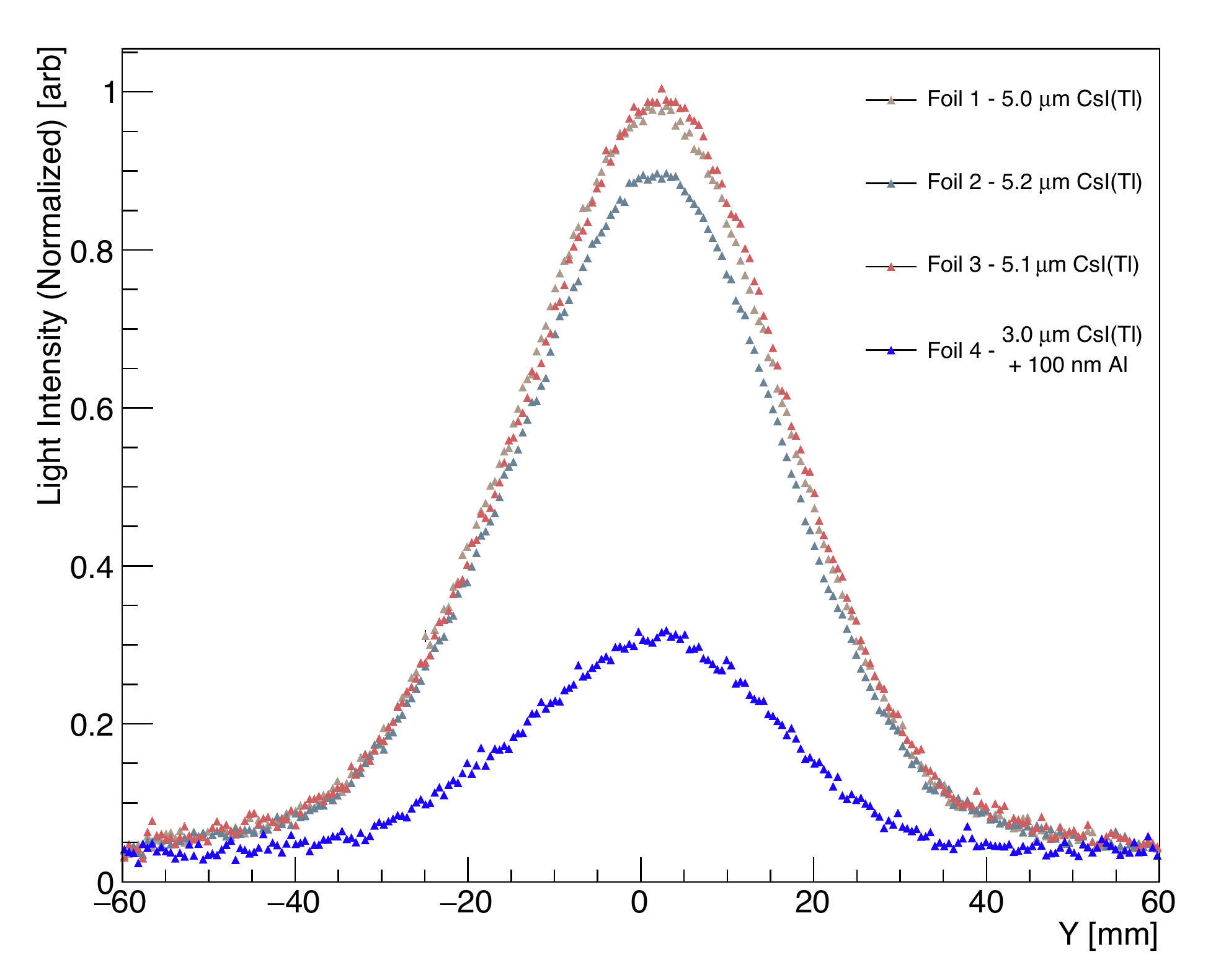}
  \label{fig:12b}
  \caption{}
  \end{subfigure}
  \caption{Shows a comparison of the four foils outlined in \Cref{tab:1}, when exposed to muons under the same conditions. (Top) shows the measured horizontal profile for all foils while (Bottom) shows the corresponding vertical profiles.}
  \label{fig:12}
\end{figure}

The first set of measurements undertaken was a comparison of the light-yield (LY) of the foils detailed in \Cref{tab:1}. 
Each foil was consecutively placed in the foil holder in the light-tight box, placed at the intermediate focus of the beam line and irradiated with the muon beam for the same exposure time. 
The foils were then analyzed as outlined in \Cref{sec:2.2} and the muon beam horizontal and vertical profile projections made (cf. \Cref{fig:12}) in order to select the best foils for further study. 
Although the measured sigmas are all comparable and within 200 $\mu$m of that measured with the pill scintillator scanner system, the light intensities differ significantly.
Foils 1 and 3 are comparable in intensity and show the highest light yield (LY). 
Foil 2 has $\sim$ 10\% lower LY at the centre though has the thickest CsI layer (0.1-0.2~$\mu$m thicker compared to foils 1 and 3), while the lowest LY is from the foil with the thinnest CsI layer and the extra Al-mirror layer. 
These measurements suggest that no quantitative inference can be made between LY and layer thickness due to the minimal differences in layer thickness and that the LY differences between foils 1-3 are most likely due to variations in production. 
The effect of the Al-mirror layer on the LY of foil 4 is not conclusive as the much lower LY is most likely due to the thickness of the CsI(Tl) layer of only 3.0~$\mu$m. 
Based on these measurements foils 1 and 3 were selected for further study.

\begin{figure}[H]
  \centering
  \begin{subfigure}[b]{0.49\columnwidth}
  \centering
  \includegraphics[width=\linewidth]{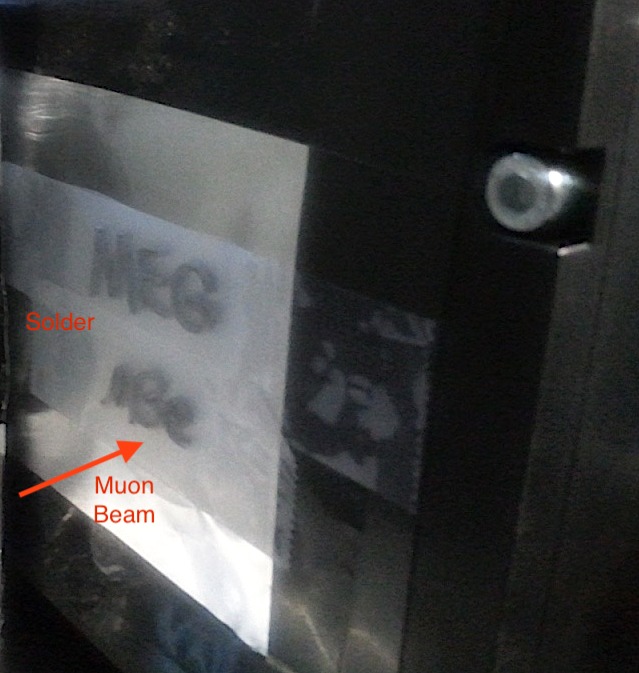}
  \caption{}
  \label{fig:13a}
  \end{subfigure}
\hfill
  \begin{subfigure}[b]{0.49\columnwidth}
  \centering
  \includegraphics[width=\linewidth]{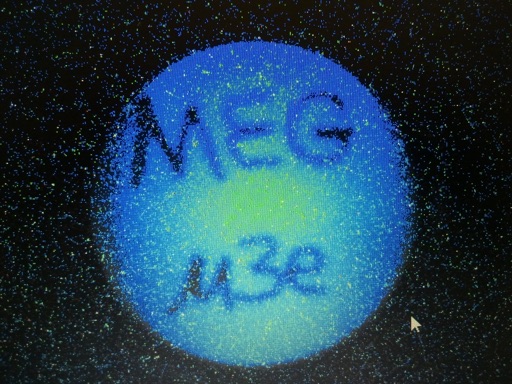}
  \caption{}
  \label{fig:13b}
  \end{subfigure}
  \caption{(Left) Muon radiograph setup showing the experimental logos made from solder-wire mounted on the exterior of the light-tight box window. (Right) shows the raw luminophore image of the logos taken with the CCD camera inside the light-tight box.}
  \label{fig:13}
\end{figure}

The next step was to characterize the resolution of the combined system of muon beam, luminophore and CCD system. 
An initial qualitative approach was first tried out by making a simple muon beam radiograph of the MEG and Mu3e experiment logos, constructed of 1 mm diameter solder-wire attached to the outside of the light-tight box as shown in \Cref{fig:13}. 
The raw CCD image of the muon beam exposed logos, in which the muons stop in the solder itself, is taken with the camera inside the light-tight box and apart from showing the beam spot itself, clearly demonstrates that the luminophore acts perfectly well in producing a muon radiograph.
A more quantitative measure of the systems spatial resolution was made by placing a 0.75~mm thick Al-grid in front of the luminophore and exposing it to the muon beam.

\begin{figure}[H]
   \centering
   \includegraphics[width=\linewidth]{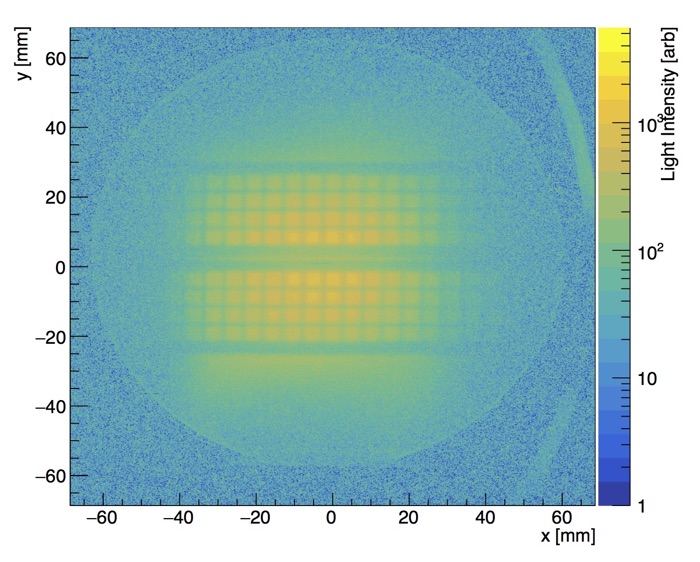}
   \caption{Muon radiograph of an aluminium grid structure used to measure the systems spatial resolution.}
   \label{fig:14}
\end{figure}

\begin{figure}[H]
  \centering
  \begin{subfigure}[t]{\columnwidth}
  \includegraphics[width=\linewidth]{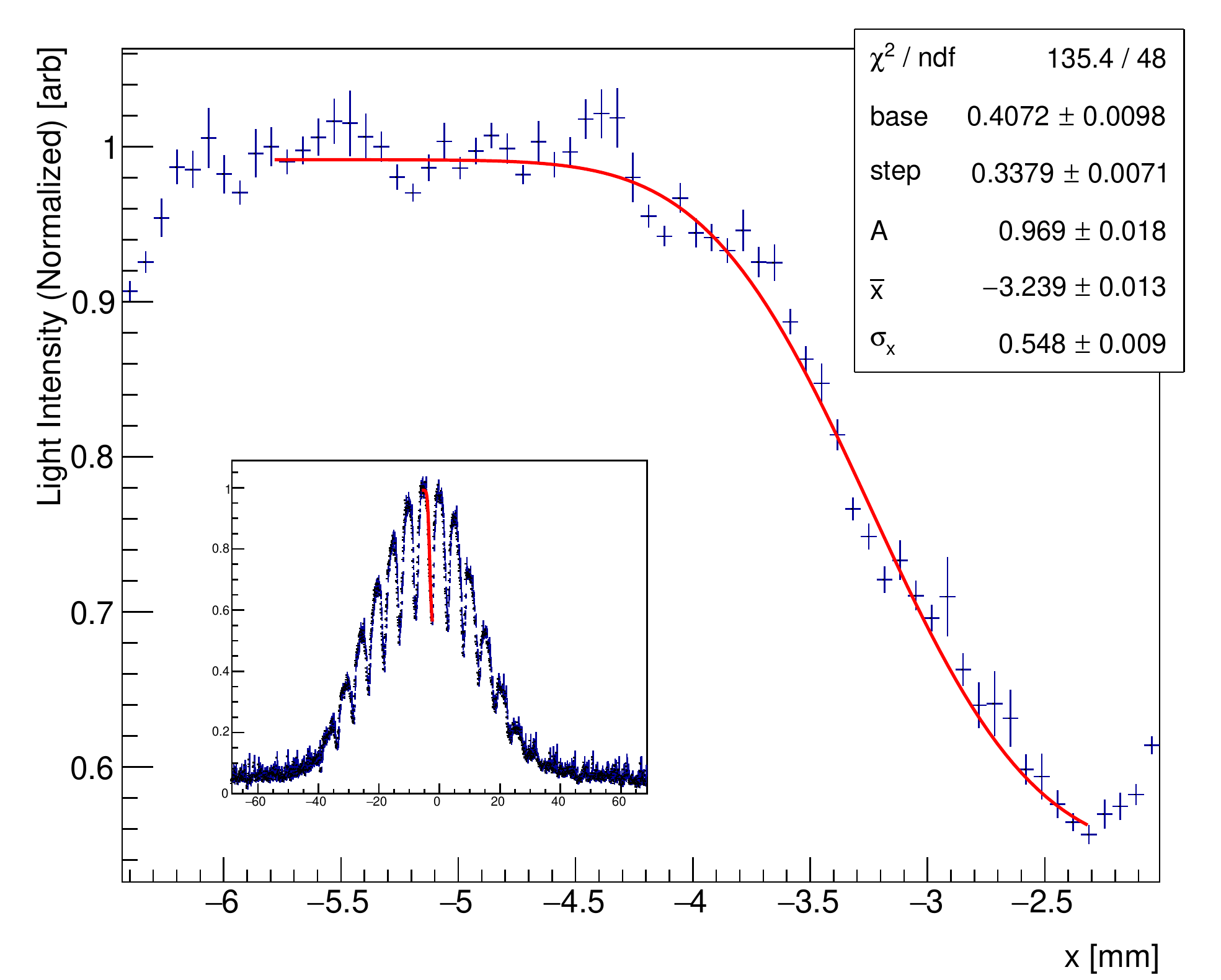}
  \caption{}
  \label{fig:15a}
  \end{subfigure}
~
  \begin{subfigure}[t]{\columnwidth}
  \includegraphics[width=\linewidth]{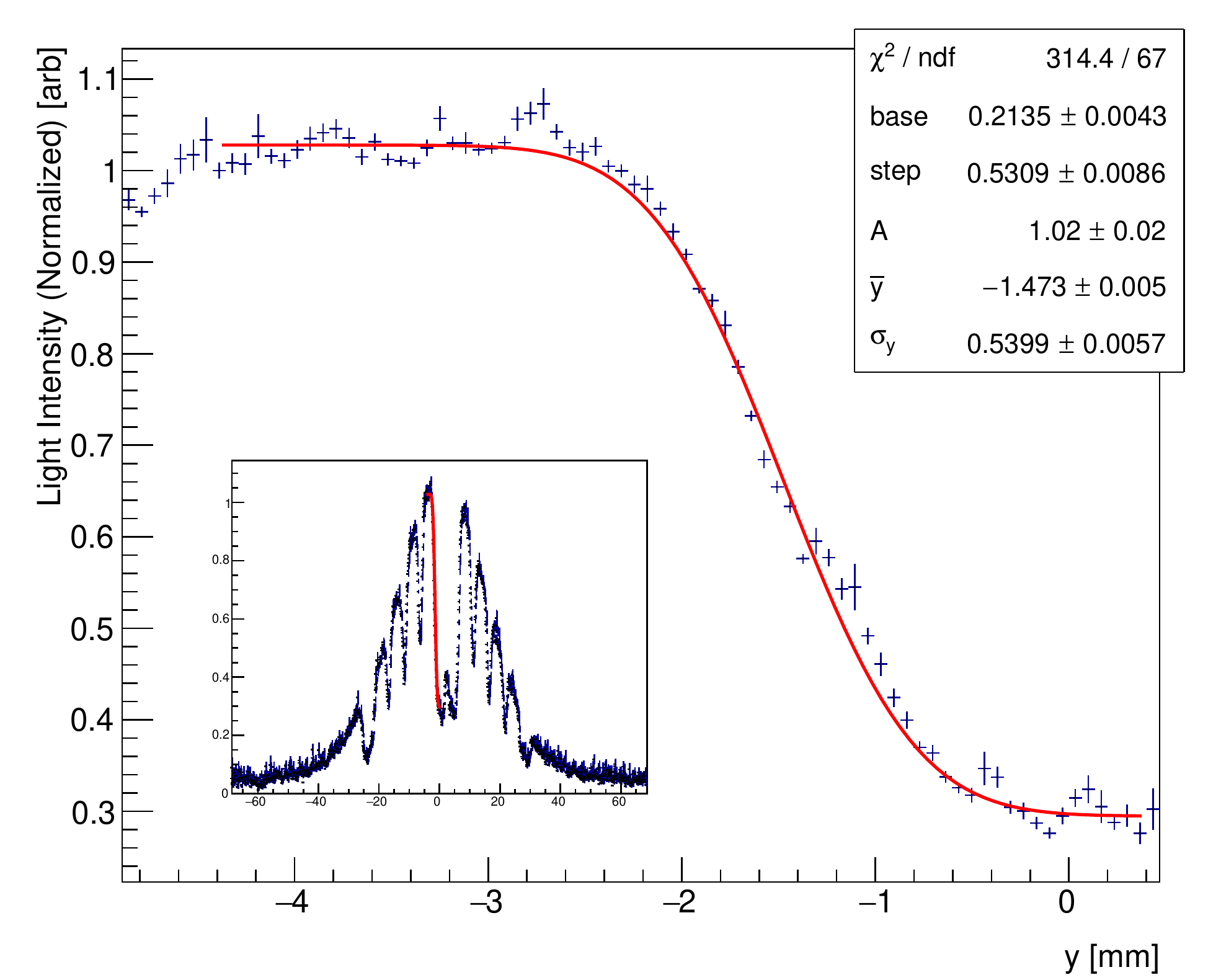}
  \caption{}
  \label{fig:15b}
  \end{subfigure}
  \caption{Luminophore radiograph of the Al-grid of \Cref{fig:14} used to determine the spatial resolution of the combined system: muon beam, luminophore \& CCD. The light intensity per pixel data points are fitted with a Gaussian convoluted step function to the edges of the openings (red curve). (Top) horizontal projection fit. (Bottom) vertical projection fit. A combined systems resolution (Beam, camera/optics \& luminophore) of 544~$\mu$m is obtained.}
  \label{fig:15}
\end{figure}

The grid is comprised of 4x4~mm$^{2}$ openings separated by 1 mm wide Al walls, a radiograph of which can be seen in \Cref{fig:14}. A horizontal and vertical projection centred on a 4x4~mm$^{2}$ gap of the grid of \Cref{fig:14} was used to fit a Gaussian convoluted step function to the edges of one of the 4x4~mm$^{2}$ opening. 
The fits provide an upper estimate on the spatial resolution of the combined system of muon beam, luminophore and CCD camera of 544~$\mu$m as shown in \Cref{fig:15}.
The inserts in the figures show the light intensity profile over the full image.
The intrinsic resolution of the luminophore itself is much higher \cite{Kozyrev2016} but the spatial resolution is dominated by the divergence of the muon beam, the muon multiple scattering and the Michel positron background from stopped muon decay in the grid adding to the LY.

\begin{figure}[H]
   \centering
   \includegraphics[width=\linewidth]{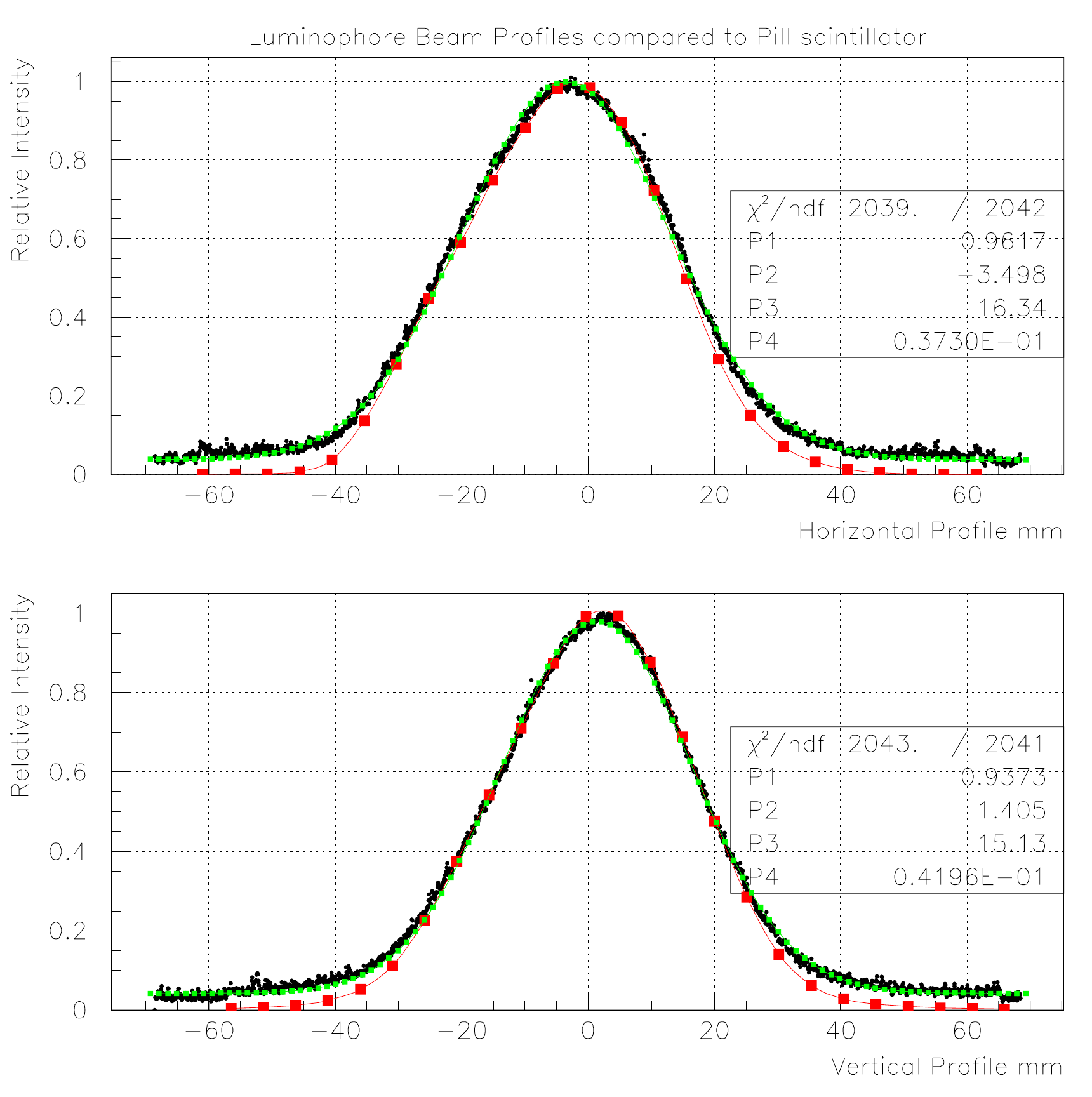}
   \caption{Comparative horizontal and vertical beam profile measurements taken with the luminophore foil (black points) and corresponding fit (green points) as compared to those of the pill scintillator scanner system (red points with fit).}
   \label{fig:16}
\end{figure}

A direct comparison of the pill scintillator system and the luminophore system was made by comparing beam profile measurements made at the same intermediate focus location, in air, just downstream of the collimator system. The individual fit values are given in \Cref{tab:3}, while the corresponding combined plots are shown in \Cref{fig:16}. 
Here the normalized measured projected vertical and horizontal luminophore intensities are shown by the black dots in the figures. 
A corresponding Gaussian plus flat background fit to these values is marked by the green dots and the fit values displayed in the figure.
The red squares show the corresponding pill scintillator profiles. 
In order to better compare the fits in the figures, the pill position data points (red squares) have been shifted slightly to coincide with the luminophore x- and y-centroids, the difference being ($\sim$ 1~mm in x and $\sim$ 3.5~mm in y) as shown by the difference in centroids in \Cref{tab:3}. 
These differences are due to a misalignment of the two beam setups. 

\begin{table}[!h]
  \centering
  \caption{Comparison of measured beam profile parameters for both the luminophore and pill scanner setups. The centroid difference is due to misalignment of the setups.}
	\begin{tabular*}{\linewidth}{@{\extracolsep{\fill}} c c c c @{}}
  	\toprule
	\thead{Beam parameter}    & \thead{Luminophore foil \\ measurement {[mm]}} & \thead{Pill scintillator \\ scanner measurement {[mm]}}  \\
	\midrule
	$\langle x \rangle$			& -3.5$\pm$0.02				& -2.56$\pm$0.09				\\
	$\boldsymbol{\sigma_{x}}$	& \bfseries 16.34$\pm$0.02	& \bfseries 16.25$\pm$0.11		\\
	$\langle y \rangle$			& 1.41$\pm$0.02				& -2.14$\pm$0.09				\\
	$\boldsymbol{\sigma_{y}}$	& \bfseries 15.13$\pm$0.02	& \bfseries 15.26$\pm$0.11		\\
	\bottomrule
 	\end{tabular*}
  \label{tab:3}
\end{table}

\begin{figure}[H]
   \centering
   \includegraphics[width=\linewidth]{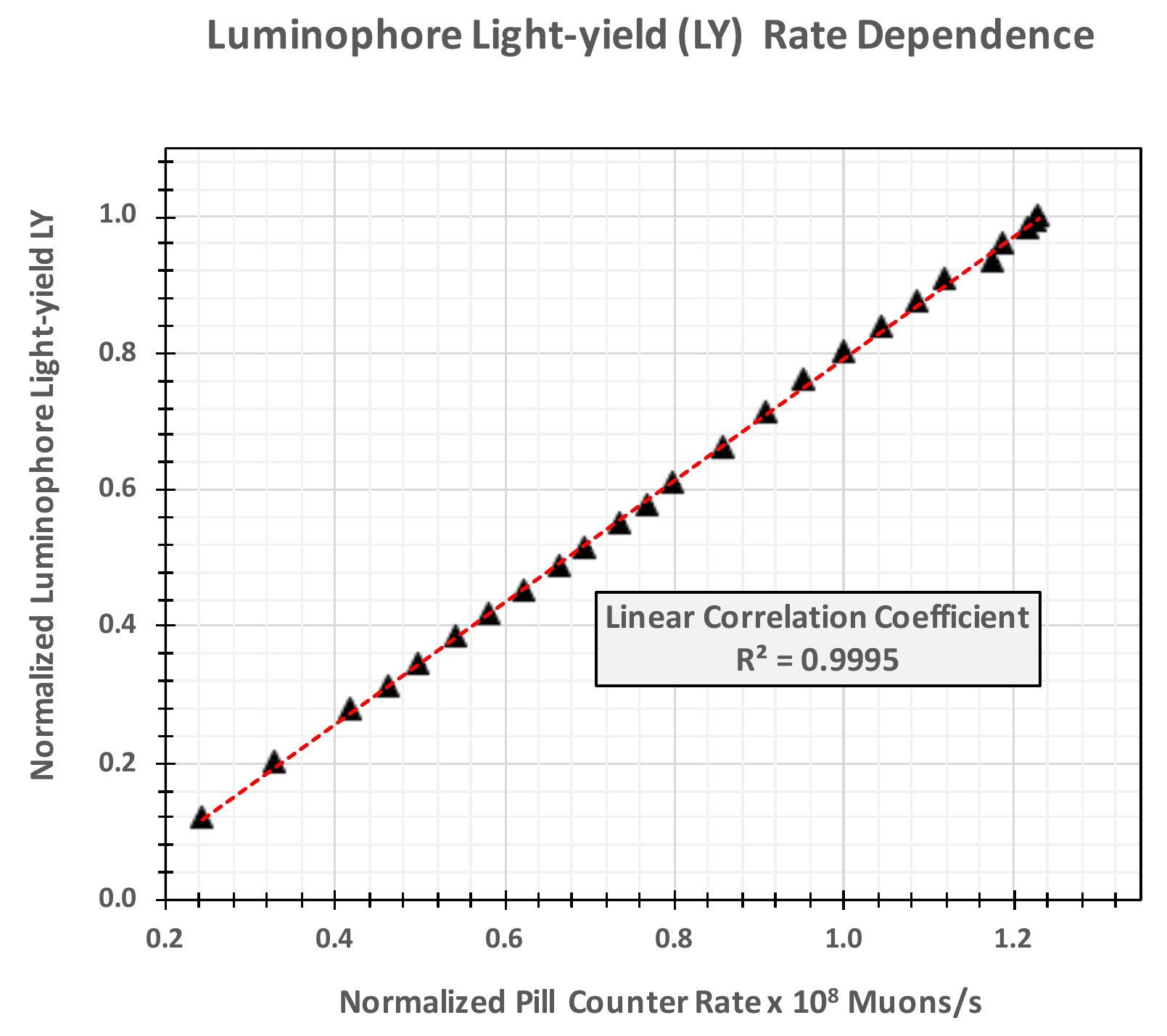}
   \caption{Luminophore linearity check of the light-yield versus muon beam intensity up to $\sim$ $1.2 \cdot 10^{8} \mu^{+}$/s. The beam intensity is varied by altering the opening of upstream slits in the channel.}
   \label{fig:17}
\end{figure}

The non-Gaussian tails exhibited by the luminophore data in \Cref{fig:16} are attributed to a flat Michel positron background caused by muons ranging-out in the air and material downstream of the foil. 
This is not present in the pill scintillator data, due to the applied pulse-height threshold for selecting muons only. 
The comparative results for the distribution widths in \Cref{fig:16} show excellent agreement ($\sim$ 100~$\mu$m) and consistency within the estimated precisions.	
The linearity of the luminophore foil versus muon beam intensity is clearly demonstrated in \Cref{fig:17}. 
Here LY-measurements with the luminophore foil were undertaken by varying the opening of the horizontal FS41L/R slits in the front-part of the channel and taking a series of beam and background images for each slit setting. 
The absolute beam intensity measurements versus slit setting were made with cross-scans using the pill-scintillator scanner system. 
A very clear linear behavior of the LY is seen over a factor of five in rate, up to the maxiumum available intensity of $\sim$ 1.2$\cdot$10$^{8}$~$\mu^{+}$/s.

A further and somewhat more involved example of the measured characteristic beam properties using the luminophore foil is the determination of the central beam momentum of a surface muon beam, in this case, at the end of the CMBL beam line. 
Here one of the most precise methods, using a detector, simultaneous multiple particle types and the accelerator radiofrequency signal as a clock to measure the time-of-flight (TOF) cannot be used due to the lack of time-structure associated with the surface muons, as well as the lack of a third particle-type (pions) at this momentum. 
Hence, we compare an integral range-curve measurement using the pill-scintillator to a corresponding range-curve measurement (Bragg-curve) with the luminophore foil to determine the beam momentum. 
The two methods are fundamentally different. 
In the case of the integral range-curve using a pill scintillator which is thicker than the range of the surface muons (2 mm), particles reaching the detector are counted versus increasing degrader thickness. 
By counting the sum of both the stopped muons and the through-going decay Michel positrons one ensures an accurate muon count is obtained with the changing muon pulse-height towards the end of its range. This can be considered as a digital method. 
The Bragg-curve, in the case of the luminophore foil is an analogue method, obtained by measuring the particle light-yield, proportional to both the energy deposition in the foil and the number of through-going particles versus degrader thickness. 

\begin{figure}[H]
   \centering
   \includegraphics[width=\linewidth]{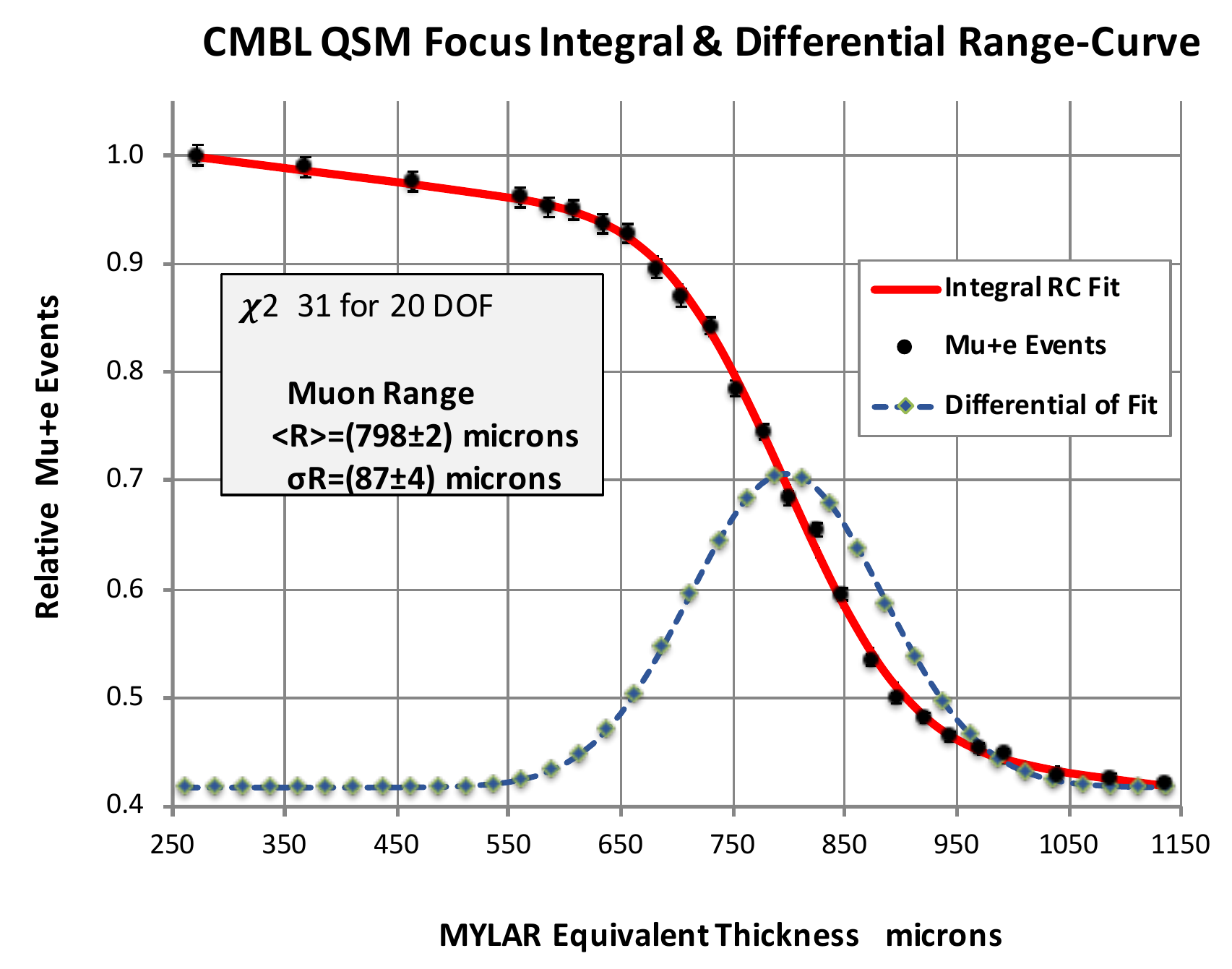}
   \caption{Integral muon range-curve measured at the final focus of CMBL beam line. The black dots correspond to the range-curve data taken with the pill scintillator. The red fitted curve represents a complimentary erf-function fit to the data points with a linear background. The green diamonds and dashed curve is a plot of the differential of the integral fit curve and serves to demonstrate that the mean range corresponds to the peak of the range-straggling distribution.}
   \label{fig:18}
\end{figure}

The integral range-curve is shown in \Cref{fig:18}. 
The black dots and associated red curve show a fit to the integral range-curve data by a complimentary erf-function with mean corresponding to the mean range $\langle$R$\rangle$ and a sigma equal to the square-root of the quadratic sum of the range-straggling $\sigma_{R}$ and the momentum-byte $\Delta$p of the channel. 
A linear background of Michel positrons is also assumed. 
For reasons of clarity and to compare to the corresponding Bragg-curve obtained by the luminophore, the differential range curve, based on the fit to the integral curve is shown by the green diamonds and dashed line in the plot. 
A mean range of 798~$\mu$m of Mylar\textsuperscript{\textregistered}~equivalent of density $\rho$=1.4 g/cm$^{3}$ is obtained, this includes the full material budget of vacuum window, air and light-tight materials. 
The total contribution from range-straggling as well as the momentum-byte yields a 1$\sigma$ value of 87~$\mu$m.

\begin{figure}[H]
   \centering
   \includegraphics[width=\linewidth]{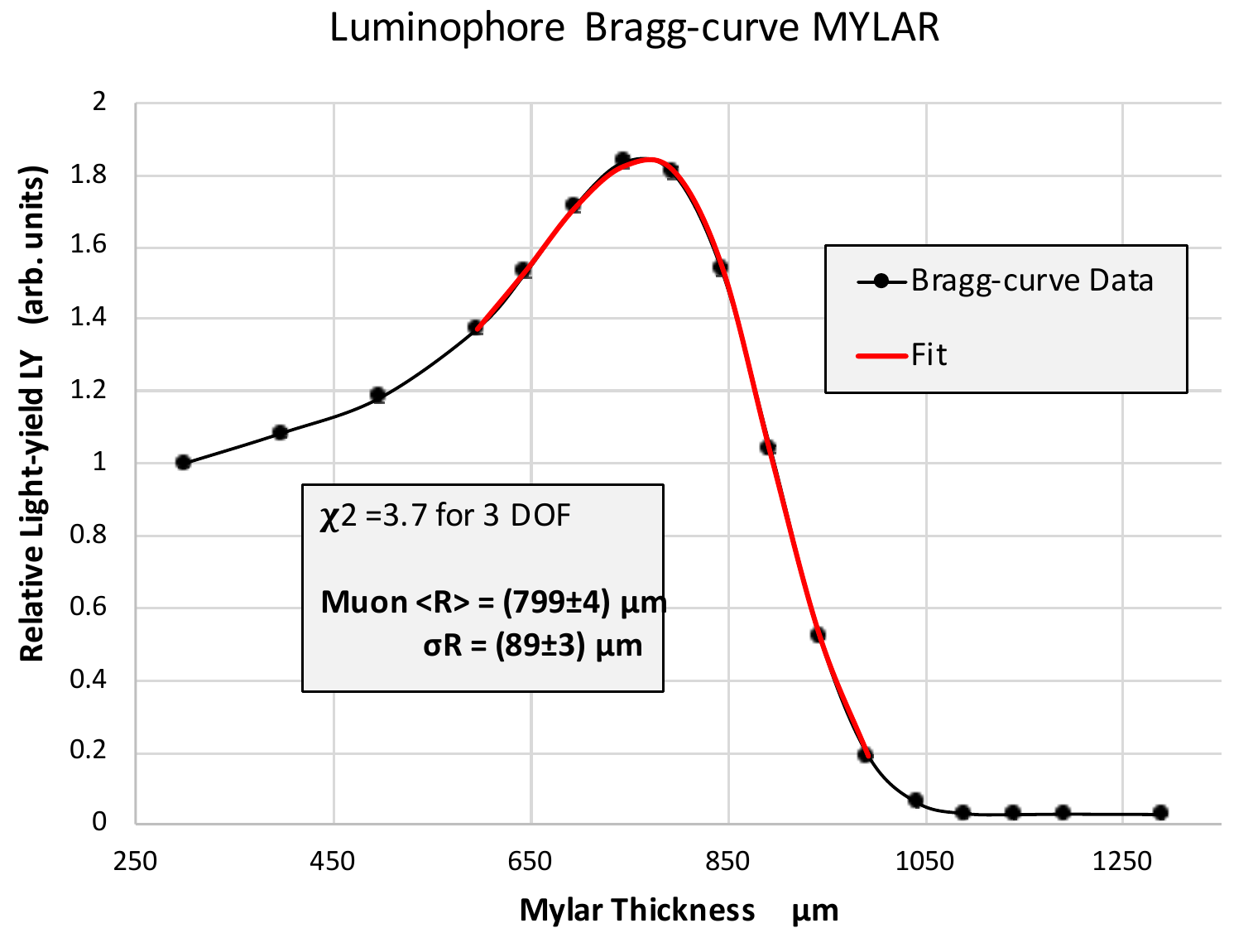}
   \caption{Shows the corresponding Bragg-curve for muons measured by the luminophore foil. The red curve shows a two-piece Gaussian plus linear background fit to the data.}
   \label{fig:19}
\end{figure}

The equivalent Bragg-curve as measured by the luminophore foil is shown in \Cref{fig:19}. 
Here the LY is normalized to the initial point corresponding to no degrader, only the material budget of the beam line and setup. 
The Bragg-curve differs from the former range-curve in so much as it being an analogue method, with the LY, based on energy deposition, being proportional to both the number of particles and their residual dE/dX after the degrader. 
The red curve is a fit to the data using a two-piece normal distribution with a linear background.
 
Both the range-curve and the Bragg-curve show excellent agreement with mean ranges and range-straggling values that differ only by a few microns, with uncertainties of less than 1\%.

The mean muon range associated with both measurements is obtained, in the case of the pill range-curve, from the thickness corresponding to the mid-way drop-off in muon rate. 
This point is equivalent to a step-function for monochromatic muons but is smeared with a Gaussian range-straggling distribution to account for energy-loss straggling of the actual muons and the finite momentum-byte of the channel.
It corresponds to the peak stopping rate, as shown by the peak of the Gaussian differential range-curve in \Cref{fig:18}. 
In the case of the Bragg-curve the mean range is given by the peak of the curve since this corresponds to the maximum energy-loss in the foil. Here, the $\delta$-like peak for a monochromatic beam is also smeared by a Gaussian straggling distribution, with a width corresponding to the same contributions as in the case of the range-curve.

The momentum is obtained from the range-momentum relationship of surface muons in Mylar\textsuperscript{\textregistered}, which corresponds to an expected power law relationship \cite{Pifer1976,Badertscher1985} with $\langle R \rangle = a \cdot P^{3.5}$ where $a$ is a material based constant. 
The range of surface muons in Mylar\textsuperscript{\textregistered}~was modeled using two simulation packages, SRIM \cite{Ziegler2010} and G4beamline \cite{Roberts201X} and the combined results are shown in \Cref{fig:20} below. 
Both simulations are consistent with a power law fit to data over the momentum range shown, and the displayed combined fit yields a momentum exponent of (3.53$\pm$0.03) and hence consistent with the expected P$^{3.5}$-behaviour from energy-loss. 
The momentum for the range-curve and the Bragg-curve are then obtained by inverting this relationship. 
The thus obtained mean central beam momentum and the individual ranges and range-straggling values are presented in \Cref{tab:4}. 
The mean central momentum of 27.30~MeV/c is obtained from the inverted range-momentum relationship using the weighted means of the individual ranges in \Cref{tab:4}. This carries a total uncertainty of 2.8\%, where the dominant contribution comes from the systematic uncertainty of the exponent in the range-momentum relationship.

\begin{figure}[H]
   \centering
   \includegraphics[width=\linewidth]{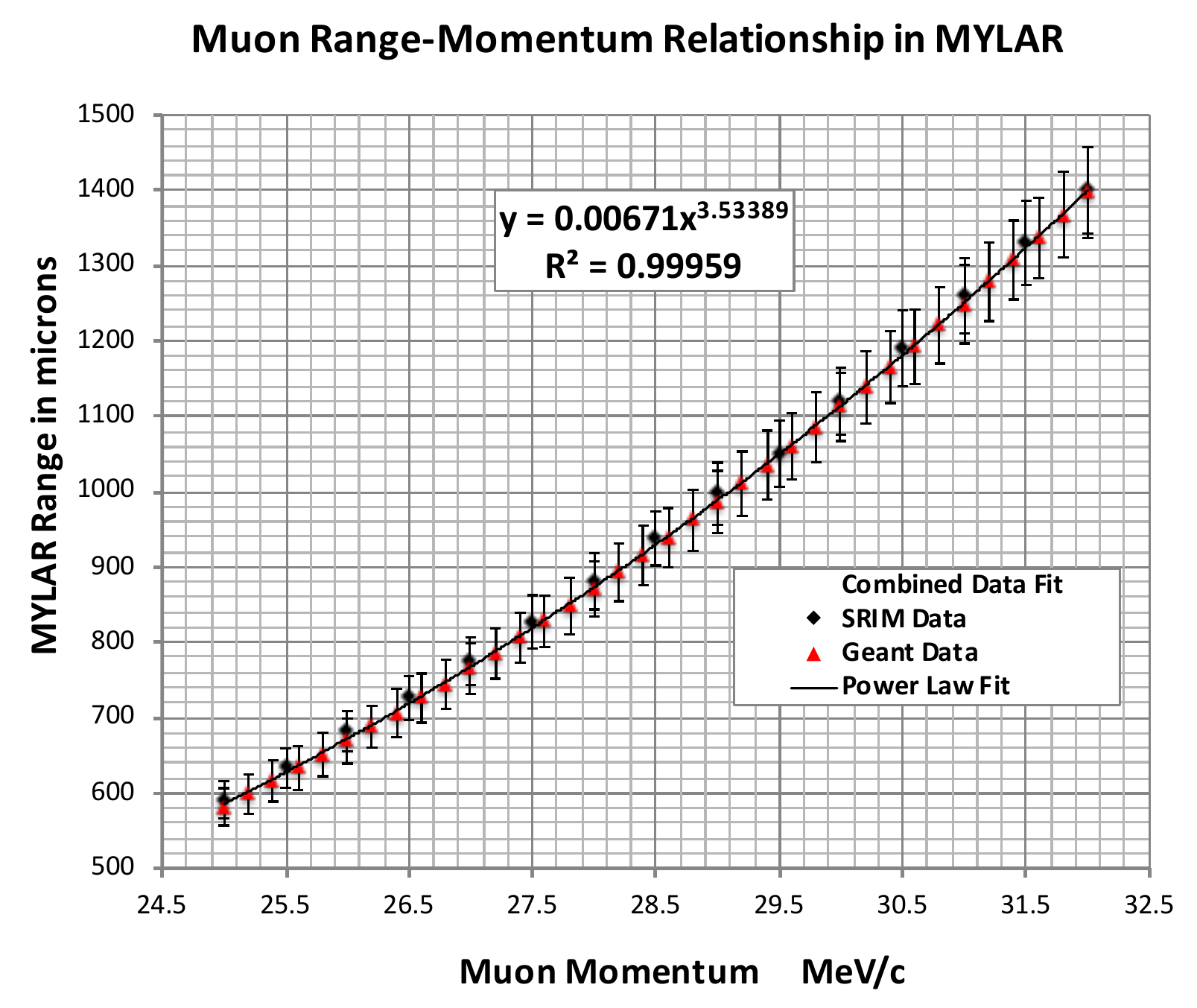}
   \caption{SRIM and G4beamline simulation data for the range versus momentum in Mylar\textsuperscript{\textregistered}~ ($\rho$=1.4 g/cm$^{3}$). The error bars represent the calculated range-straggling. A power law fit to the data gives an exponent of 3.53$\pm$0.03.}
   \label{fig:20}
\end{figure}

\begin{table}[!h]
  \centering
  \caption{Comparison of the two range methods showing the individual fit values for the mean ranges and range-straggling as well as the mean central beam momentum and momentum-byte including range-straggling obtained from the range-momentum relationship.}
	\begin{tabular*}{\linewidth}{@{\extracolsep{\fill}} c c c c @{}}
  	\toprule
	\thead{Beam parameter}    & \thead{Range-curve \\ measurement } & \thead{Bragg-curve \\ measurement}  \\
	\midrule
	$\langle R \rangle$ [$\mu$m]	& 798$\pm$2		& 799$\pm$4	\\
	$\sigma_{R} $ [$\mu$m]			& 87$\pm$4		& 89$\pm$3	\\
	\midrule

	\multicolumn{3}{c}{Mean Values}  \\
	\midrule
	$\langle P_{0} \rangle$	[MeV/c]	& \multicolumn{2}{c}{27.30$\pm$0.75}	\\
	$\sigma_{\Delta P} $	[MeV/c]	& \multicolumn{2}{c}{0.86$\pm$0.03}	\\
	\bottomrule
	\end{tabular*}
  \label{tab:4}
\end{table}

\section{Future Prospects}\label{sec:4}
\begin{figure}[H]
   \centering
   \includegraphics[width=0.925\linewidth]{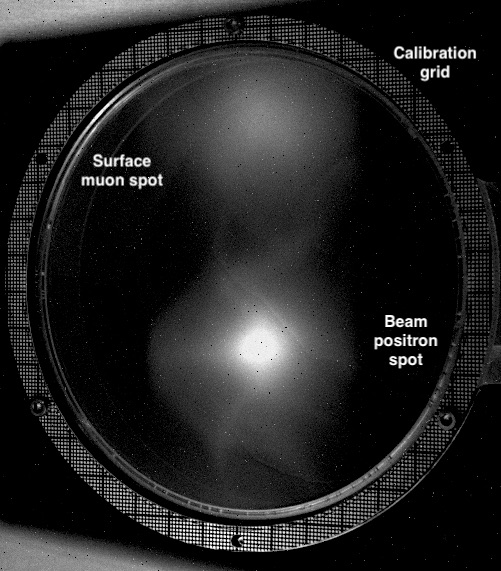}
   \caption{Luminophore foil in vacuum, introduced into the beam at the intermediate collimator system. The foil and frame are illuminated by a UV-LED in order to show both the calibration grid and support, as well as the beam spots being shown. Upper spot is the surface muon beam and the lower, more intense spot, from the beam positrons originating from the production target. The separating power of the Wien-filter has here purposely been reduced to see both beam spots.}
   \label{fig:21}
\end{figure}

Finally, with a study for a new high-intensity surface muon beam underway at PSI (HiMB Project), based on a new high-power target design \cite{Berg2016} and solenoidal channel, which would allow next-generation cLFV-experiments as well as $\mu$SR-experiments to access muon rates at the 10$^{10}$-level per second and so require new beam monitoring possibilities. 
We are, in this context, also investigating the use of luminophore foils integrated into an automated beam monitoring system for high intensities. 
With an avalanche photodiode detector (APD) placed at a fixed position in the beam spot tails, just in front of the luminophore, an automatic intensity normalization for the luminophore could be obtained. 
As a first probe of the intensity criterion we subjected foil~1, placed in vacuum at the intermediate collimator position (see \Cref{fig:10}), to simultaneous beams of muons and an order of magnitude more intense beam positron beam stemming from the production target. 
In order to show the capabilities of the luminophore we reduced the separation power of the Wien-filter to allow both beam spots to be imaged simultaneously on the foil but spatially separated. \Cref{fig:21} show the CCD image of the two spots as well as the holder and calibration grid, made possible by illuminating the system with a UV-LED. 
The upper spot is the surface muon beam, while the lower, more intense spot, is from beam positrons. 
The summed beam intensity shown here is $\sim$ 10$^{9}$~particles/s. 
Although the beam positrons can clearly be distinguished from the muons here, on account of the spatial separation, it has to be remembered that the luminophore detector does not have any particle identification capabilities as the LY is essentially proportional to the product of particle rate and the energy-loss in the foil. 
Further studies of the linearity of the LY with intensity are required if this method is to be successfully used for secondary beam intensities of the order of 10$^{10}$~particles/s.

\section{Conclusions}\label{sec:5}
We have developed a non-invasive, ultra-thin, CsI(Tl) luminophore foil detector system, based on CCD-imaging that has successfully been implemented into an intensity-frontier surface muon beam at PSI. 
The system is still under development in order to achieve its automation for beam optimization and experiment monitoring. 
We have outlined the production and the image analysis and have demonstrated the usefulness of such a non-destructive method. 
Intrinsic optical and mechanical properties, relevant to high-intensity secondary particle beams, have also been presented. 
Fundamental beam characteristics have been measured and compared to our standard, `destructive' beam methods of measurement and show excellent agreement at the sub-millimetre level, as well as the total systems resolution of the luminophore system which is also at this level. 
Furthermore, the new system shows an excellent linear LY-response to beam intensity over a wide range, though this must be further studied if one is to go to an order-of-magnitude increase in particle rates. 

We have also demonstrated the complementarity of such a method in determining fundamental beam properties for precision experiments, such as beam momentum and range-straggling. 
Even though the luminophore foil system does not have particle identification capabilities, we have shown the wide range of capabilities that such a non-destructive system can contribute to the current generation of cLFV precision experiments such as those of MEG~II and Mu3e.

\section*{Acknowledgements}
We gratefully acknowledge discussions with Andreas Knecht, as well as the support of the infrastructure groups of our institutes and especially those of PSI. 
This work was also partly supported by the Swiss SNF Grant 20021$\_$137738.

%
%
\bibliography{refs}

\begin{thebibliography}{10}
\expandafter\ifx\csname url\endcsname\relax
  \def\url#1{\texttt{#1}}\fi
\expandafter\ifx\csname urlprefix\endcsname\relax\def\urlprefix{URL }\fi
\expandafter\ifx\csname href\endcsname\relax
  \def\href#1#2{#2} \def\path#1{#1}\fi

\bibitem{Crookes1903}
W.~Crookes, {Certain properties of the emanations of radium}, Chemical News
  87~(2269) (1903) 241 (1903).

\bibitem{Geiger1913}
D.~H. Geiger, E.~Marsden, {The laws of deflexion of a particles through large
  angles}, The London, Edinburgh, and Dublin Philosophical Magazine and Journal
  of Science 25~(148) (1913) 604--623 (1913).
\newblock \href {https://doi.org/10.1080/14786440408634197}
  {\path{doi:10.1080/14786440408634197}}.

\bibitem{PSIHandbook}
{PSI Users' Guide Accelerator Facilities}, {Paul Scherrer Institute} (1994).

\bibitem{Baldini2018}
A.~M. Baldini, et~al., {The design of the MEG II experiment}, The European
  Physical Journal C 78~(5) (2018) 380 (2018).
\newblock \href {https://doi.org/10.1140/epjc/s10052-018-5845-6}
  {\path{doi:10.1140/epjc/s10052-018-5845-6}}.

\bibitem{Blondel2013}
A.~Blondel, et~al., {Research Proposal for an Experiment to Search for the
  Decay $\mu \to eee$} (2013).
\newblock \href {http://arxiv.org/abs/1301.6113} {\path{arXiv:1301.6113}}.

\bibitem{Nagarkar1998}
V.~V. Nagarkar, et~al., {Structured CsI(Tl) scintillators for X-ray imaging
  applications}, IEEE Transactions on Nuclear Science 45~(3) (1998) 492--496
  (June 1998).
\newblock \href {https://doi.org/10.1109/23.682433}
  {\path{doi:10.1109/23.682433}}.

\bibitem{Cha2009}
B.~K. Cha, et~al., {Scintillation characteristics and imaging performance of
  CsI:Tl thin films for X-ray imaging applications}, Nuclear Instruments and
  Methods in Physics Research Section A: Accelerators, Spectrometers, Detectors
  and Associated Equipment 604~(1) (2009) 224 -- 228, pSD8 (2009).
\newblock \href {https://doi.org/10.1016/j.nima.2009.01.177}
  {\path{doi:10.1016/j.nima.2009.01.177}}.

\bibitem{Cosentino2003}
L.~Cosentino, P.~Finocchiaro, {Ion beam imaging at very low energy and
  intensity}, Nuclear Instruments and Methods in Physics Research Section B:
  Beam Interactions with Materials and Atoms 211~(3) (2003) 443 -- 446 (2003).
\newblock \href {https://doi.org/10.1016/S0168-583X(03)01816-0}
  {\path{doi:10.1016/S0168-583X(03)01816-0}}.

\bibitem{Harasimowicz2010}
J.~Harasimowicz, et~al., {Scintillating screens sensitivity and resolution
  studies for low energy, low intensity beam diagnostics}, Review of Scientific
  Instruments 81~(10) (2010) 103302 (2010).
\newblock \href {https://doi.org/10.1063/1.3488123}
  {\path{doi:10.1063/1.3488123}}.

\bibitem{Re2005}
M.~Re, et~al., {Production of inorganic thin scintillating films for ion beam
  monitoring devices}, Conf. Proc. C0505161 (2005) 808 (2005).

\bibitem{Eaton1994}
G.~H. Eaton, et~al., {Fast E field switching of a pulsed surface muon beam: The
  Commissioning of the European muon facility at ISIS}, Nucl. Instrum. Meth.
  A342 (1994) 319--331 (1994).
\newblock \href {https://doi.org/10.1016/0168-9002(94)90257-7}
  {\path{doi:10.1016/0168-9002(94)90257-7}}.

\bibitem{Lord2010}
J.~Lord, \href{{http://nmi3.eu/files/beam_camera_report.pdf}}{{The new Muon
  Beam Camera}}, {Muon JRA Reports} (2010).
\newline\urlprefix\url{{http://nmi3.eu/files/beam_camera_report.pdf}}

\bibitem{Ito2014}
T.~U. Ito, et~al., {Online full two-dimensional imaging of pulsed muon beams at
  J-PARC MUSE using a gated image intensifier}, Nucl. Instrum. Meth. A754
  (2014) 1--9 (2014).
\newblock \href {https://doi.org/10.1016/j.nima.2014.04.014}
  {\path{doi:10.1016/j.nima.2014.04.014}}.

\bibitem{Stoykov2005}
A.~Stoykov, et~al., {A scintillating fiber detector for muon beam profile
  measurements in high magnetic fields}, Nuclear Instruments and Methods in
  Physics Research Section A: Accelerators, Spectrometers, Detectors and
  Associated Equipment 550~(1) (2005) 212 -- 216 (2005).
\newblock \href {https://doi.org/10.1016/j.nima.2005.04.089}
  {\path{doi:10.1016/j.nima.2005.04.089}}.

\bibitem{Strasser2010}
P.~Strasser, et~al., {J-{PARC} decay muon channel construction status}, Journal
  of Physics: Conference Series 225 (2010) 012050 (apr 2010).
\newblock \href {https://doi.org/10.1088/1742-6596/225/1/012050}
  {\path{doi:10.1088/1742-6596/225/1/012050}}.

\bibitem{Ripiccini2016}
E.~Ripiccini, A.~Papa, G.~Rutar, High granularity scintillating fiber trackers
  based on silicon photomultiplier, 2016, p. 009 (2016).
\newblock \href {https://doi.org/10.22323/1.252.0009}
  {\path{doi:10.22323/1.252.0009}}.

\bibitem{SGCsI}
{Saint-Gobain CsI(Tl) ref},
  \url{https://www.crystals.saint-gobain.com/products/csitl-cesium-iodide-thallium}.

\bibitem{Kozyrev2016}
E.~Kozyrev, et~al., {Performance and Characterization of CsI:Tl thin Films for
  X-ray Imaging Application}, Physics Procedia 84 (2016) 245 -- 251 (2016).
\newblock \href {https://doi.org/10.1016/j.phpro.2016.11.042}
  {\path{doi:10.1016/j.phpro.2016.11.042}}.

\bibitem{IDS}
{Image Development Systems (IDS)}, {IDS UI-2220SE CCD camera},
  \url{https://www.1stvision.com/cameras/models/IDS-Imaging/UI-2220SE}.

\bibitem{ORCA}
{Hamamatsu}, {ORCA Flash 4.0 V2},
  \url{https://www.hamamatsu.com/resources/pdf/sys/SCAS
  0080E_C11440-22CU_tec.pdf}.

\bibitem{QSI}
{Quantum Scientific Imaging (QSI)}, {RS-9.2s CCD Camera},
  \url{http://www.qsimaging.com/rs92-overview.html}.

\bibitem{Bradski2008}
G.~Bradski, \href{{https://opencv.org}}{{The OpenCV Library}}, Dr. Dobb's
  Journal of Software Tools (2000).
\newline\urlprefix\url{{https://opencv.org}}

\bibitem{Gridin2014}
S.~S. Gridin, et~al., {Channels of Energy Losses and Relaxation in CsI:A
  Scintillators ({$\rm A$}={$\rm Tl$},In)}, IEEE Transactions on Nuclear
  Science 61~(1) (2014) 246--251 (2014).
\newblock \href {https://doi.org/10.1109/TNS.2013.2283316}
  {\path{doi:10.1109/TNS.2013.2283316}}.

\bibitem{Lagu1961}
R.~G. Lagu, B.~V. Thosar, {Fluorescence and scintillation spectra of CsI (Tl)
  crystal}, Proceedings of the Indian Academy of Sciences - Section A 53~(5)
  (1961) 219--226 (1961).
\newblock \href {https://doi.org/10.1007/BF03049186}
  {\path{doi:10.1007/BF03049186}}.

\bibitem{Ritt2010}
S.~Ritt, R.~Dinapoli, U.~Hartmann, {Application of the DRS chip for fast
  waveform digitizing}, Nucl. Instrum. Meth. A623 (2010) 486--488 (2010).
\newblock \href {https://doi.org/10.1016/j.nima.2010.03.045}
  {\path{doi:10.1016/j.nima.2010.03.045}}.

\bibitem{Elliott2007}
C.~Elliott, et~al., {National Instruments LabVIEW: A Programming Environment
  for Laboratory Automation and Measurement}, JALA: Journal of the Association
  for Laboratory Automation 12~(1) (2007) 17--24 (2007).
\newblock \href {https://doi.org/10.1016/j.jala.2006.07.012}
  {\path{doi:10.1016/j.jala.2006.07.012}}.

\bibitem{Pifer1976}
A.~Pifer, T.~Bowen, K.~Kendall, {A high stopping density $\mu^{+}$ beam},
  Nuclear Instruments and Methods 135~(1) (1976) 39 -- 46 (1976).
\newblock \href {https://doi.org/10.1016/0029-554X(76)90823-5}
  {\path{doi:10.1016/0029-554X(76)90823-5}}.

\bibitem{Badertscher1985}
A.~Badertscher, et~al., {Development of ``subsurface” positive muon beam at
  LAMPF}, Nuclear Instruments and Methods in Physics Research Section A:
  Accelerators, Spectrometers, Detectors and Associated Equipment 238~(2)
  (1985) 200 -- 205 (1985).
\newblock \href {https://doi.org/10.1016/0168-9002(85)90455-3}
  {\path{doi:10.1016/0168-9002(85)90455-3}}.

\bibitem{Ziegler2010}
J.~F. Ziegler, M.~Ziegler, J.~Biersack, {SRIM – The stopping and range of
  ions in matter}, Nuclear Instruments and Methods in Physics Research Section
  B: Beam Interactions with Materials and Atoms 268~(11) (2010) 1818 -- 1823,
  19th International Conference on Ion Beam Analysis (2010).
\newblock \href {https://doi.org/10.1016/j.nimb.2010.02.091}
  {\path{doi:10.1016/j.nimb.2010.02.091}}.

\bibitem{Roberts201X}
T.~Roberts, \href{{http://G4beamline.muonsinc.com}}{G4beamline package}.
\newline\urlprefix\url{{http://G4beamline.muonsinc.com}}

\bibitem{Berg2016}
F.~Berg, et~al., {Target Studies for Surface Muon Production}, Phys. Rev.
  Accel. Beams 19~(2) (2016) 024701 (2016).
\newblock \href {https://doi.org/10.1103/PhysRevAccelBeams.19.024701}
  {\path{doi:10.1103/PhysRevAccelBeams.19.024701}}.

\end{thebibliography}

\end{document}